# Controlling Orbital Ordering of Intergrowth Structures with Flat [$Ag^{(II)}F_2$] Layers to Mimic Oxocuprates(II)†

Daniel Jezierski,*[a] Jose Lorenzana,*[b] and Wojciech Grochala*[a]

Based on the Density Functional Theory calculations, we propose a new pathway toward compounds featuring flat [$AgF_2$] layers which mimic [$CuO_2$] layers in high-temperature oxocuprate superconductor precursors. Calculations predict the dynamic (phonon) and energetic stability of the new phases over diverse substrates. For some compounds with ferro orbital ordering, we find a gigantic intrasheet superexchange constant of up to –211 meV (DFT+U) and –256 meV (SCAN), calculated for hypothetical $(CsMgF_3)_2KAgF_3$ intergrowth. Semiempirical calculations show that at optimum doping, the expected superconducting critical temperature should reach 200 K. The partial substitution of $K^+$ with $Ba^{2+}$ leads to noticeable electron doping of [$AgF_2$] sublattice, as revealed by progressive population of the Upper-Hubbard band. On the other hand, modest 10-15% hole-doping through partial substitution of $Mg^{2+}$ with $Li^+$, primarily leads to the depopulation of $p_{(z)}$ orbitals of apical F atoms. We also find structures with an undesired antiferrodistortive structural ordering and discuss the structural factors that determine the transition from buckled to flat planes and from different types of orbital ordering using Landau theory of phase transitions.

## Introduction

The record of the superconducting critical temperature ($T_c$) of around 130K in the $Hg_{0.8}Tl_{0.2}Ba_2Ca_2Cu_3O_{8+\delta}$ phase[1] at ambient pressure, was set shortly after Bednorz and Müller's discovery of high temperature superconductivity (HTSC) in Ba-doped $La_2CuO_4$ ($T_c$ = 30K)[2]. Since then, many studies have been devoted to new HTSC analogs in the hope of achieving even higher $T_c$ than the existing ones. This led to the discovery of superconductivity in iron pnictides with a record $T_c \approx 56K$[3], nickelates with a record $T_c \approx 17K$[4], and, according to the latest findings, cobaltite with $T_c$ of about 6K[5]. Moreover, FeSe thin film on a $SrTiO_3$ substrate achieved through the use of molecular beam epitaxy (MBE) has a $T_c$ near 100K[6], which is tenfold higher than for its bulk form[7]. However, this is still far from the long-sought goal of room-temperature superconductivity at ambient pressure.

Although the nature of HTSC phenomenon remains largely a mystery, there are several specific properties that cuprates precursors exhibit: (i) strong antiferromagnetic superexchange ($|J_{2D}|$)[8,9], which can occur in systems with flat layers[10–12], (ii) 2D topology of interaction[13], and (iii) the charge-transfer (CT) nature of the fundamental gap[14]. Consequently, a significant number of materials have been recently proposed[15–17], among which Ag-F systems[18] and especially the layered $Ag^{II}F_2$[19,20], have particularly gained significant attention.

Semiempirical considerations show that $T_c$ in cuprates scales with the magnitude of antiferromagnetic interactions[21,22]. Large antiferromagnetic exchange requires a ferro orbital[23]. In contrast, layered ternary compounds, with alkali metals, $M^I_2Ag^{II}F_4$ (M=Na-Cs), may seem to be promising precursors for HTSC. However, an antiferro orbital order leads to an antiferrodistortive (AFD) arrangement of the [$AgF_2$] sheets which reduces their potential as HTSCs precursors[24–26]. Related M-$Ag^{II}$-F systems, $M^IAg^{II}F_3$ (M=K-Cs), feature undesired, low-dimensional 1D antiferromagnetic ordering[27–29]. For alkaline earth and closed shell transition metals systems of $M^{II}Ag^{II}F_4$ formula (M=Ba, Sr, Ca, Cd, Hg)[25,30], the presence of isolated flat squares [$AgF_4$]$^{2-}$ in their structures, excludes the possibility of strong antiferromagnetic coupling in 2D. The only known $M_2Ag^{II}F_6$ system (M=Ba) is paramagnetic[30], exhibiting $Ag^{2+}$ centers with square planar configuration[18].

These considerations show that, in addition to avoid or reduce the puckering of the layers it is essential to control their orbital ordering. Consequently, attention is focused on the simplest, binary compound – $Ag^{II}F_2$. Despite the pronounced puckering of the [$AgF_2$] layers[31], the intrasheet superexchange constant for $AgF_2$ is around –70meV[32], where we follow the convention that negative interactions correspond to antiferromagnetism. This constitutes half of the value characteristic to cuprates[33]. Furthermore, this binary fluoride shares several other properties with HTSC precursors: CT nature of the gap[34,35], a pronounced covalent character of metal-ligand bond[36] and a striking similarity in electronic structure as revealed by hybrid density of states (HSE06) calculations[32]. Nevertheless, in its bulk form, the pronounced inclination towards the localization of extra charges may result in formation of self-trapped states[37,38]. Therefore, theoretical investigations have led to the identification of two potential solutions to this issue. Our very recent work identified the possibility to stabilize a new polytype of $AgF_2$, distinguished by its enhanced doping propensity and higher $|J_{2D}|$, comparing to the ambient form[39]. Alternatively, the stabilization of an almost flat monolayer of $AgF_2$ on an appropriate substrate has been proposed[21]. This method also allows for doping[37,40,41], potentially assuring the $T_c$ at around 200K[21] for silver(II) difluoride in the monolayer form.

[a] d.jezierski@cent.uw.pl, w.grochala@cent.uw.edu.pl
Center of New Technologies, University of Warsaw, 02089 Warsaw, Poland
[b] jose.lorenzana@cnr.it
Institute for Complex Systems (ISC), CNR, 00185 Rome, Italy
† This work is dedicated to Prof. Grzegorz Chałasiński at his 75th birthday.
Supplementary Information available: structures and magnetic properties, energetics of formation reactions, density of states, and doping to Cs2K and LCO systems. See DOI: 10.1039/x0xx00000x

In this paper, we introduce the third approach to utilize the potential of the $[AgF_2]$ layers – via formation of the intergrowth structures. Similar types of structure are frequently found for cuprate high-$T_C$ precursors[42] and many become high-$T_C$ superconductors upon doping[43]. In this study we use theoretical calculations to assess stability and properties of diverse intergrowth systems, corresponding to various compositions of $MAgF_3$, $M_2AgF_4$, $MgF_2$ and $AgF_2$ substrates (M=K, Rb, Cs). We find that the transition from buckled to planar structures and from ferro to antiferro orbital ordering is driven by the planar strain and construct a Landau theory which allows to draw a generalized phase diagram of the $AgF_2$ layer.

## Computational methods

Density Functional Theory (DFT) calculations were performed, implemented in the VASP 5.4.4 code[44]; GGA with PBEsol[45] functional and projected augmented wave method was used[46,47]. DFT+U with Liechtenstein rotationally invariant[48] was introduced for on-site Coulombic interactions of open-shell *d* electrons; U = 5 eV and J = 1 eV for Ag[39,49,50] – providing good agreement with experimental geometry parameters. The values of intra-sheet superexchange constants, $J_{2D}$, was also calculated using HSE06[51] and SCAN[52] methods. A plane-wave cutoff energy of 520 eV was used in all systems. The k-mesh was set to 0.032Å$^{-1}$ for geometry optimization and 0.022Å$^{-1}$ for self-consistent-field convergence. The convergence thresholds of $10^{-7}$ for ionic and $10^{-9}$eV for electronic steps were used. Phonon curves were calculated employing PHONOPY package[53].

To assess the potential for charge doping into the $d_{(x2-y2)}$ orbital manifold of silver, we carried out calculations using the Virtual Crystal Approximation (VCA) method[54]. This approach enables the simulation of gradual cation substitution by generating a virtual pseudopotential as a weighted average of the components involved at the same site. Crucially, it allows for the simulation of the substitution without the need for large supercells with reduced symmetry (which would otherwise be required for individual substitution at specific sites), thereby avoiding substantial computational costs. We have studied two types of isostructural substitution, both well-known from the literature: $K^+ \rightarrow Ba^{2+}$ and $Mg^{2+} \rightarrow Li^+$.

It should be noted, that $K_2AgF_4$ and $Rb_2AgF_4$ can form solid solutions with complete miscibility of K/Rb cations[24]. Therefore, the additional possibility of forming intergrowth systems which are disordered at alkali metal site, cannot be excluded. However, in this preliminary study we have decided to keep the K-Rb-Cs sublattice fully ordered to minimize computational cost. Similar assumption applies to a possible Ag/Mg disorder.

## Results and discussion

### A Structural analysis

Theoretical DFT calculations permit these days to screen possible crystal structures and properties of compounds before they are prepared in the laboratories. Such approach is less energy- and resource-consuming and it permits selection of the most promising systems.

Here, we constructed and optimized (DFT+U) hypothetical intergrowth structures in either orthorhombic (*Pnma*) or monoclinic ($P2_1/c$) unit cells, where $[AgF_2]$ layers are found between $MF/MgF_2$ sublattices (Figure 1). The intergrowth compound is composed of stacked sheets arranged as $[AgF_2][MF][MgF_2][MF][MgF_2][MF]$. Variation of M (M=K, Rb, Cs) allows to modify strain between sublattices and influence the geometry within the $[AgF_2]$ sheet. In other words, these bulk systems may show beneficial features similarly to those predicted theoretically when a single $AgF_2$ sheet is placed on a substrate with appropriate lattice constant[21]. Typically, the distance between $[AgF_2]$ layers is approximately 13 Å, ensuring very weak inter-sheet interactions.

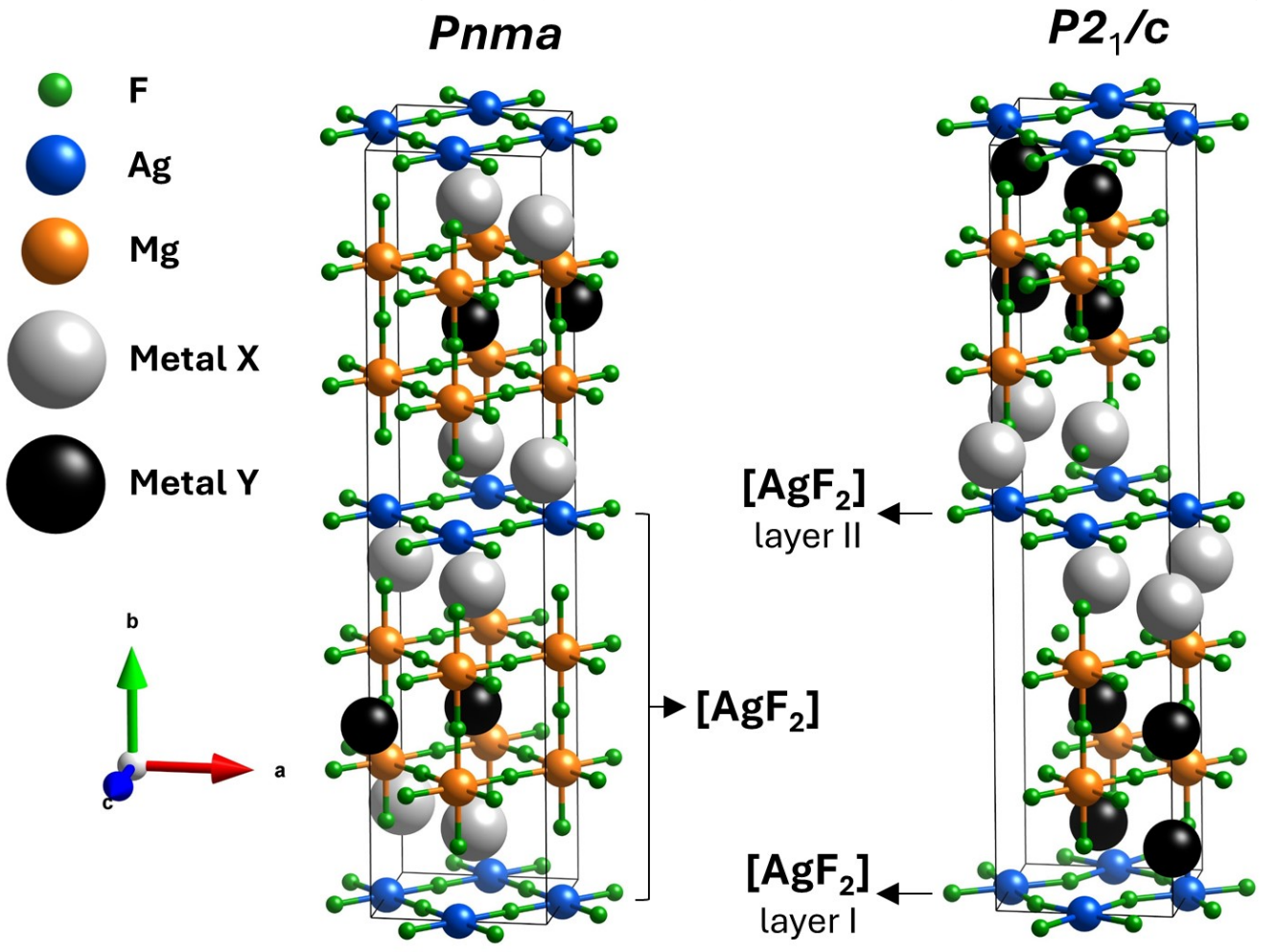

Figure 1. Structures adopted for calculations. Color code: F-green, Ag-blue, Mg-orange, grey and black for two different metal ions from group I of the periodic table. For monoclinic $P2_1/c$ space group, three different alkali metal ions were also considered (e.g. RbCsK structures).

The *Pnma*-type structures are characterized by crystallographically identical $[AgF_2]$ sheets; the alkali metal always forms two identical MF layers. In the monoclinic structure ($P2_1/c$), however, the distinction between two different $[AgF_2]$ layers becomes apparent, as a different alkali metal (X/Y) may neighbor each of them. Therefore, these $[AgF_2]$ layers are denoted as layer I and layer II for clarity. In the following analysis, we will focus on the structural aspects of the $[AgF_2]$ layers after full geometric optimization of the intergrowth systems. Particularly, we will analyze the puckering and occurrence of antiferrodistortive ligand arrangement, which are crucial factors influencing the strength and nature of magnetic intrasheet coupling between paramagnetic $Ag^{2+}$ ions.

Table 1. Key structural parameters for optimized orthorhombic *Pnma* intergrowth structures. **A**, **C**, **α** correspond to specific parameters of the $[AgF_2]$ layer, adhering to the notation introduced in Figure 2. **B1/B2** is a measure of the intra-layer distortion. For precise values check Table SI1 in SI. Metal X denotes the metal cation in proximity to the silver difluoride layer as depicted in Figure 1, thereby forming the [MF] layer. $AgF_2$, α-$K_2AgF_4$, $Rb_2AgF_4$ and $Cs_2AgF_4$ were optimized as references.

| Metal X | Short name | Compound | A [Å] | C [Å] | α [°] | B1/B2 |
|---|---|---|---|---|---|---|
| 2xCs | Cs3 | $(CsMgF_3)_2CsAgF_3$ | 4.18 | 2.49 | 179.8 | 1.00 |
| 2xCs | Cs2Rb | $(CsMgF_3)_2RbAgF_3$ | 4.15 | 2.51 | 179.8 | 1.00 |
| 2xRb | CsRb2 | $(RbMgF_3)_2CsAgF_3$ | 4.14 | 2.36 | 175.8 | 0.95 |
| 2xCs | Cs2K | $(CsMgF_3)_2KAgF_3$ | 4.13 | 2.52 | 179.6 | 1.00 |
| 2xK | CsK2 | $(KMgF_3)_2CsAgF_3$ | 4.12 | 2.22 | 148.0 | 0.89 |
| 2xRb | Rb3 | $(RbMgF_3)_2RbAgF_3$ | 4.08 | 2.42 | 163.6 | 1.00 |
| 2xRb | Rb2K | $(RbMgF_3)_2KAgF_3$ | 4.05 | 2.44 | 159.2 | 1.00 |
| 2xK | RbK2 | $(KMgF_3)_2RbAgF_3$ | 4.02 | 2.39 | 151.0 | 1.00 |
| 2xK | K3 | $(KMgF_3)_2KAgF_3$ | 3.99 | 2.40 | 149.4 | 1.00 |
| --- | --- | $AgF_2$ | 3.73 | 2.57 | 128.7 | 1.00 |
| K | | α-$K_2AgF_4$ (*Bmab*)* | 4.46 | 2.10 | 159.8 | 0.84 |
| Rb | --- | $Rb_2AgF_4$ | 4.44 | 2.10 | 179.7 | 0.86 |
| Cs | --- | $Cs_2AgF_4$ | 4.54 | 2.10 | 180.0 | 0.83 |

The optimized intergrowth systems exhibit Ag-Ag distances ranging from 3.99 to 4.18 Å (Table 1). These values are larger than the one for $AgF_2$ with **A**=3.73 Å (Figure 2); in which the $[AgF_2]$ layers display significant puckering, with an **α** angle equal to 128.7°. Additionally, **A** parameters in the hypothetical structures are smaller than those for $Rb_2AgF_4$ (4.44 Å) and $Cs_2AgF_4$ (4.54 Å), where there is almost no puckering, yet the distortion of layer is marked – as indicated by the coefficient **B1/B2** < 1 (**B1** and **B2** state for lengths of equatorial bonds). Hence, it can be inferred that the layer puckers when the distances between Ag-Ag are too short, and it distorts when these distances are excessively long. This trend was also theoretically predicted in case of $[AgF_2]$ monolayer deposited on diverse substrates[21].

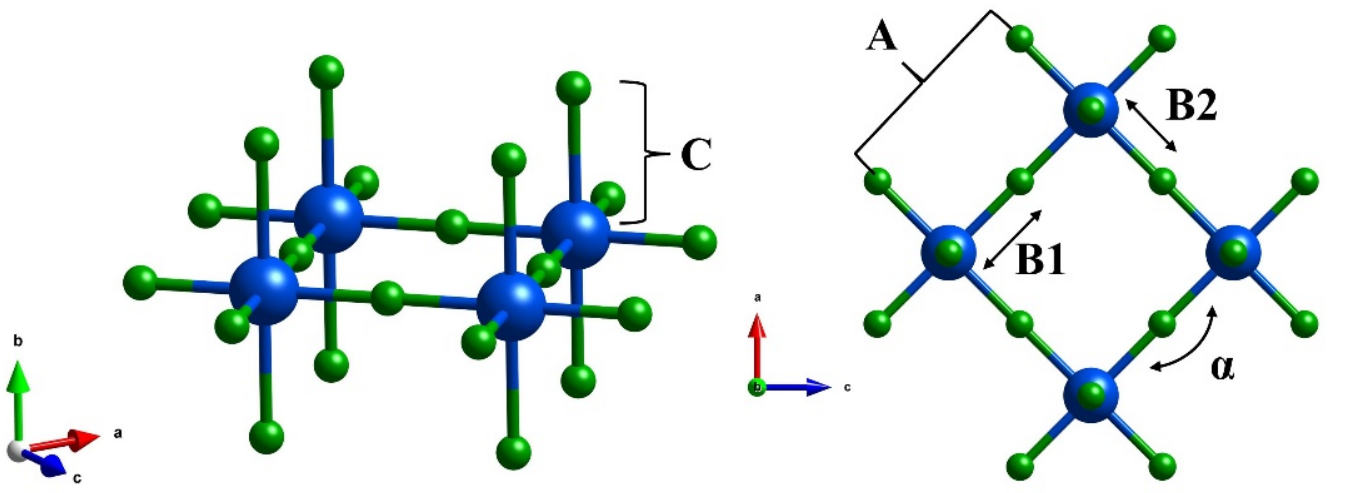

Figure 2. $[AgF_{4/2+2/1}]$ layer in intergrowth structure, with indicated bond lengths **C** - axial Ag-F, **B1** and **B2** - equatorial Ag-F bonds, Ag-Ag distances as **A**, and **α** for the Ag-F-Ag bond angle in the layer.

A correlation between the F-Ag-F angle (**α**) and the Ag-Ag distance (**A**) seen previously for $[AgF_2]$ monolayer[21] is also present in intergrowth systems. In the system with the smallest lattice constant within the studied series, designated as K3 (Table 1), the most pronounced puckering of the layer is observed, where **α**=149.4°. Conversely, for the system labeled as Cs3, which has the largest value of **A** (4.18Å), the $[AgF_2]$ layer is nearly flat (179.8°). In both systems, there is no distortion of the silver difluoride sublattice (**B1/B2**=1). However, In the cases of CsK2 (**A**=4.12 Å) and CsRb2 (**A**=4.14 Å), an antiferrodistortive arrangement of the $[AgF_2]$ layer is observed. Since chemical pressure is intermediate between the Cs3 and K3 ones[24] this suggests that other factors besides chemical pressure determines the fate of the $[AgF_2]$ layer. We will come back to this point below.

Let us now examine results for the $P2_1/c$ systems (Table 2). The key distinction in structural properties between $P2_1/c$ and *Pnma* lies in the presence of two distinct $[AgF_2]$ layers, arising from their proximity to two different [MF] sheets composed of two distinct alkali metals (as metal X or Y from Figure 1). The monoclinic symmetry also enables the exploration of systems containing three different cation ions in the structure ($Rb^+$, $Cs^+$, $K^+$) simultaneously – denoted as mI_RbCsK, mII_RbCsK, and mIII_RbCsK (Table 2).

Different ionic environments of the two $[AgF_2]$ layers result in their distinct structures. However, the general features seen for *Pnma* systems are similar to those observed for $P2_1/c$ structures. For most systems, the $[AgF_2]$ layers show no Ag-F bond alternation. However, in three cases (mCs2Rb, mII_RbCsK, mRb2K) the antiferrodistortive pattern appears. As we show below, the above results may be rationalized by performing a Landau analysis of the structural transitions.

Table 2. Key structural parameters for monoclinic $P2_1/c$ intergrowth structures. **A**, and **C** correspond to the individual types of bonds within the $[AgF_2]$ layer, following the notation presented in Figure 2. **B1/B2** is a measure of the intra-layer distortion. For precise values *cf.* Table SI2 in ESI. Metal Y, X corresponds to the metal cation in direct proximity to the silver difluoride layer – following the notation from Figure 1. Metal Y forms a layer adjacent to layer I of $AgF_2$ (labeled as I), while metal X is adjacent to layer II (labeled as II).

| Metal Y , X | Short name | Compound | A [Å] | C [Å] | | α [°] | | B1/B2 | |
|---|---|---|---|---|---|---|---|---|---|
| | | | | *I* | *II* | *I* | *II* | *I* | *II* |
| Cs, Rb | mCs2Rb | $(CsMgF_3)_2RbAgF_3$ | 4.17 | 2.51 | 2.32 | 179.9 | 173.8 | 1.00 | 0.93 |
| Rb, Cs | mCsRb2 | $(RbMgF_3)_2CsAgF_3$ | 4.12 | 2.40 | 2.55 | 169.5 | 179.9 | 1.00 | 1.00 |
| Rb, K | mI_RbCsK | $RbCsKMg_2AgF_9$ | 4.11 | 2.40 | 2.26 | 167.0 | 150.4 | 0.99 | 0.90 |
| K, Cs | mII_RbCsK | $RbCsKMg_2AgF_9$ | 4.11 | 2.24 | 2.58 | 149.9 | 179.9 | 0.90 | 1.00 |
| Cs, Rb | mIII_RbCsK | $RbCsKMg_2AgF_9$ | 4.09 | 2.57 | 2.42 | 179.6 | 164.3 | 1.00 | 1.00 |
| K, Cs | mCsK2 | $(KMgF_3)_2CsAgF_3$ | 4.08 | 2.28 | 2.59 | 149.7 | 179.6 | 0.91 | 1.00 |
| Rb, K | mRb2K | $(RbMgF_3)_2KAgF_3$ | 4.06 | 2.44 | 2.32 | 160.8 | 145.0 | 1.00 | 0.92 |
| K, Rb | mRbK2 | $(KMgF_3)_2RbAgF_3$ | 4.02 | 2.39 | 2.47 | 151.3 | 156.6 | 1.00 | 1.00 |

**B Landau analysis of structural transitions in strained $AgF_2$ layers**

Intuitively, one can expect that a flat $AgF_2$ layer subject to compressive strain will buckle. Furthermore, previous work has shown that tensile strain induces an antiferrodistortive state with bond ordering and a staggered ordering of $e_g$ orbitals. These are typical symmetry lowering transitions that can be studied using Landau theory of phase transitions. If we assume that the geometry of the $[AgF_2]$ layer is determined solely by the chemical strain imposed by the outer layers we can construct a generalized Landau free energy which is valid across different systems. Deviation from the predictions of this theory will indicate other factors other than chemical stain influencing the structure.

To construct the Landau free energy of the $[AgF_2]$ layer we start by defining the different order parameters. We define an order parameter to describe the buckling as $\boldsymbol{\phi} = \cos\left(\frac{\alpha}{2}\right) = \frac{\boldsymbol{h}}{\boldsymbol{d_{AgF}}}$, where $\boldsymbol{h}$ is the distance of the fluorine to the Ag-Ag and $\boldsymbol{d_{Ag-F}}$ the Ag-F distance. $\boldsymbol{\phi = 0}$ characterizes the flat plane while $\boldsymbol{\phi \neq 0}$ reprsents a buckled plane. Likewise, we can define an order parameter for the bond-order AFD transition as $\boldsymbol{\psi} = \frac{\boldsymbol{B2-B1}}{\boldsymbol{B2+B1}} = \frac{\boldsymbol{1-B1/B2}}{\boldsymbol{1+B1/B2}}$ where $\boldsymbol{\psi \neq 0}$ represents a plane with bond order. We can use $\boldsymbol{A}$ as the control parameter (strain). For small values of the order parameters, the energy difference between the low-symmetry phase and the high symmetry phase can be expanded in powers of the order parameters. Assuming in the high symmetry phase the layer coincides with a mirror plane and Ag sites have $C_4$ symmetry, only even powers of the order parameters can be present, $\boldsymbol{\delta F = \alpha\,\phi^2 + \beta\,\psi^2 + \gamma\,\phi^2\,\psi^2 + \eta\phi^4 + \zeta\,\psi^4}$, where $\boldsymbol{\alpha}$, $\boldsymbol{\beta}$, $\boldsymbol{\gamma}$, $\boldsymbol{\eta}$, and $\boldsymbol{\zeta}$ are constants in the order parameters but may depend on external parameters like strain, temperature, and the characteristic of the system (elastic modulus, etc.). According to Landau theory, to describe a continuous phase transition one should have the leading behavior $\boldsymbol{\alpha = \alpha_1(A_{c1} - A)}$ and $\boldsymbol{\beta = \beta_1\,(A - A_{c2})}$ with $\boldsymbol{\alpha_1, \beta_1}$ also constants and $\boldsymbol{\eta, \zeta > 0}$. In this way, when $\boldsymbol{\psi = 0}$ for $\boldsymbol{A > A_{c1}}$ the free energy is minimum for $\boldsymbol{\phi = 0}$ (flat plane) while for $\boldsymbol{A < A_{c1}}$ the energy is minimized by $\boldsymbol{\phi \neq 0}$ (buckled plane) as schematically shown in Figure 3.

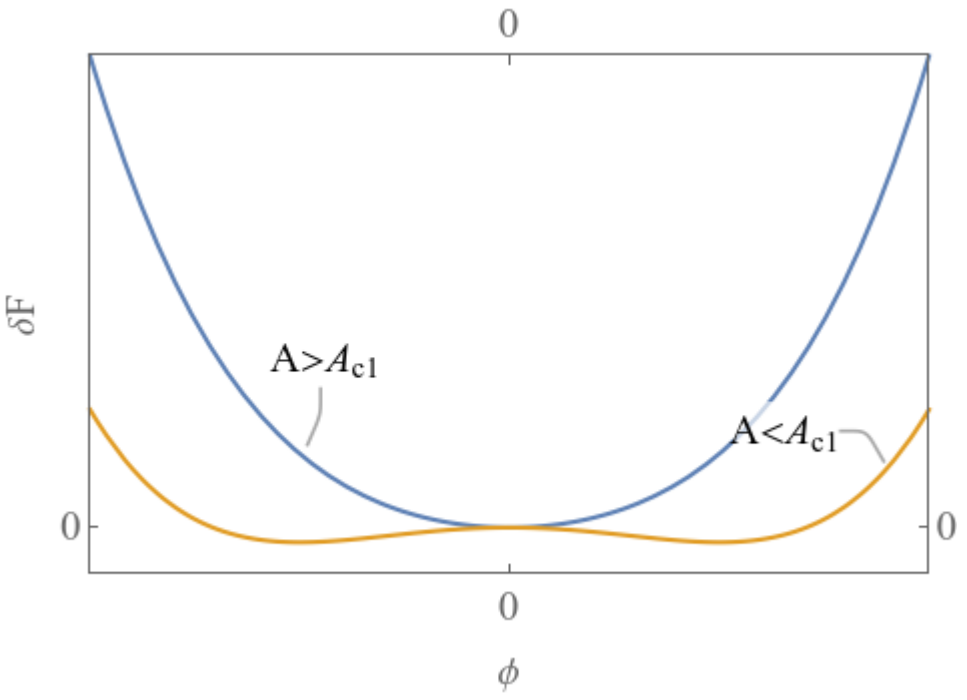

Figure 3. Schematic Landau free energy for the case in which $\psi = 0$ describing the transition from a flat plane ($A > A_{c1}$) to a buckled plane ($A < A_{c1}$). Analogously, setting $\psi = 0$, for $A < A_{c2}$ all planar AgF bond are equal while for $A > A_{c2}$ the bonds disproportionate with the AF distortive pattern.

If the ordering transitions are determined mainly by the characteristics of the $AgF_2$ plane, then we expect that the constants in the Landau free energy to depend weakly on the kind of outer planes. According to our assumptions, the main role of the outer planes is to provide the strain and enter though the value of $\boldsymbol{A}$. Landau theory predicts for a continuous transition in the absence of coexistence the order parameter behavior, $\boldsymbol{\phi = \sqrt{\alpha_1(A_{c1} - A)/(2\eta)}}$ and $\boldsymbol{\psi = \sqrt{\beta_1\,(A - A_{c2})/(2\zeta)}}$. Figure 4 shows a collection of data from this work for intergrowth phases and from previous work for strained monolayers on different substrates and form bulk (non-intergrowth systems[37]). We see that for a variety of systems $\boldsymbol{\phi^2}$ is indeed approximately linear with $\boldsymbol{(A_{c1} - A\;)}$ for $\boldsymbol{A < A_{c1}}$ until the critical point defined by $\boldsymbol{A_{c1} = 4.13}$ Å. Surprisingly, bulk $AgF_2$ also follows in this "universal" line (see inset) although Landau expansion is well justified only close to the critical point. For $\boldsymbol{\psi}$ we do not have too many systems in the "critical" region, but an approximate linear behavior can be obtained with $\boldsymbol{A_{c2} \approx 4.25}$ Å.

Some outliers can be identified. $AgF_2$ layers for CsK2, mCsK2 I, mRbCsK (I and II) and mRb2K II show coexistence of both order parameters. In a second order scenario, this is controlled by the parameter $\boldsymbol{\gamma}$ with $\boldsymbol{\gamma < 0}$ favoring phase coexistence. In the present case, however, the coexisting points are far from the critical region which suggests a first order scenario. This requires more terms in the Landau expansion and goes beyond our present scope. Another group of outliers (RbMgF2, KZnF3,

mCsK2 II, mIIIRbCsK, $SnF_4$, mIRbCsK I, mCsRb2 II) is given by anomalously flat systems with $A < A_{c1}$. This suggests that the transition between flat and buckled is actually weakly first order (i.e. first order with a small barrier) and that outer planes details can tip the balance between the two phases in the critical region. In this case the above order parameter behavior is still valid if extended to the metastable region. Interestingly, the non-intergrowth compounds $K_2AgF_4$, $Rb_2AgF_4$, and $Cs_2AgF_4$ are close to the universal layer behavior except that $K_2AgF_4$ shows substantial buckling although it has $A > A_{c1}$.

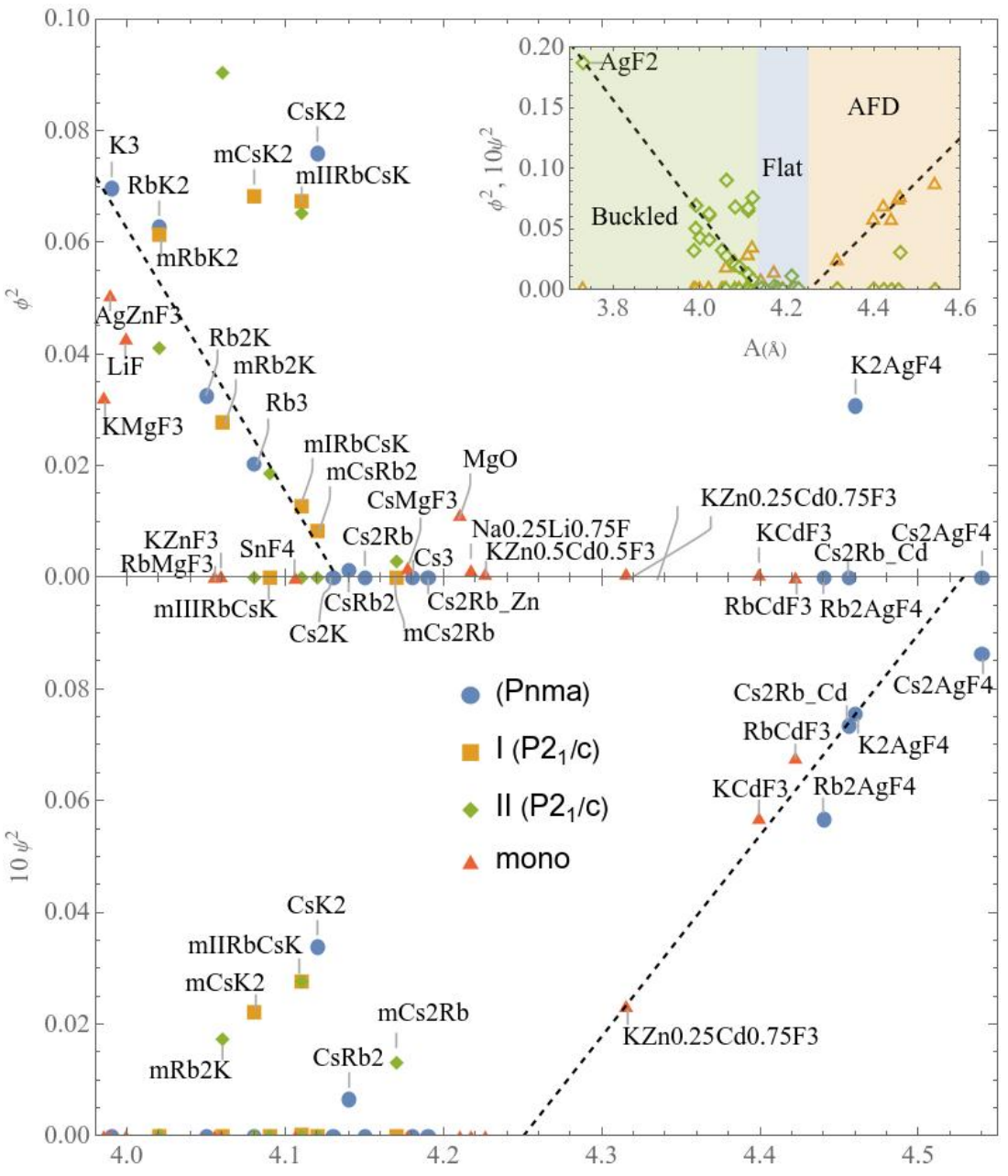

Figure 4. squared order parameter describing the buckling transition (upper panel) and the antiferrodistorsive transition (lower panel). We show present work results for *Pnma* (blue dots), $P2_1c$ in layer I (orange squares) and layer II (green diamonds) and from previous work on monolayers (red triangles). For the present work we use the same labels as in the Tables 1,2 while monolayers are labeled by the substrate. For the $P2_1c$ structure labeled data correspond to layer I while the corresponding layer II is the green diamond with the same value of A. Also labels for small values of $\psi^2$ are omitted but can be deduced from the labels in the upper panel with the same A. Inset: Zoom out of the data with the resulting phase diagram ($\phi^2$: green diamonds, $10\psi^2$: open orange triangles). Notice that the AFD is also flat but with the unwanted orbital order.

Clearly for the outlayers, strain alone does not explain the geometry and other interactions with the outer layers have to be taken into account. On the other hand, a large number of systems are well explained by Landau theory. In this case, $AgF_2$ layers in different chemical environments shows substantial universal behavior in the sense that the fate of the layer depends weakly on the details of the environment and is mainly determined by the equal biaxial strain described by the parameter A. In systems with $A > A_{c1}$ the buckled state is unstable. For systems with $A < A_{c1}$ a buckled state is favored but a flat state is also possible depending on details. For $A < A_{c2}$ the AFD state is unstable while for $A > A_{c2}$ one finds the AFD order. This provides a nearly universal phase diagram (inset of Figure 4) which should facilitate the search for ideal structures with flat planes and no AFD.

**C Magnetic and electronic properties**

For all intergrowth structures we calculated $J_{2D}$ – the magnetic intrasheet superexchange constant (SE) between $Ag^{2+}$ cations in the Ag/F layer. The $J_{2D}$ values for those hypothetical structures and $AgF_2$ are presented in the Table SI1-2 in ESI. Positive value of $J_{2D}$ refers to ferromagnetic whereas negative to antiferromagnetic ordering between spins. By stronger SE we always mean a larger absolute value of $J_{2D}$, i.e. $|J_{2D}|$. In the Figure 5 we present dependence between Ag-F equatorial and apical bond lengths and $J_{2D}$ value for *Pnma* structures. These two structural parameters appear to have the strongest influence on the $J_{2D}$ values (see Figures SI3-4 in ESI).

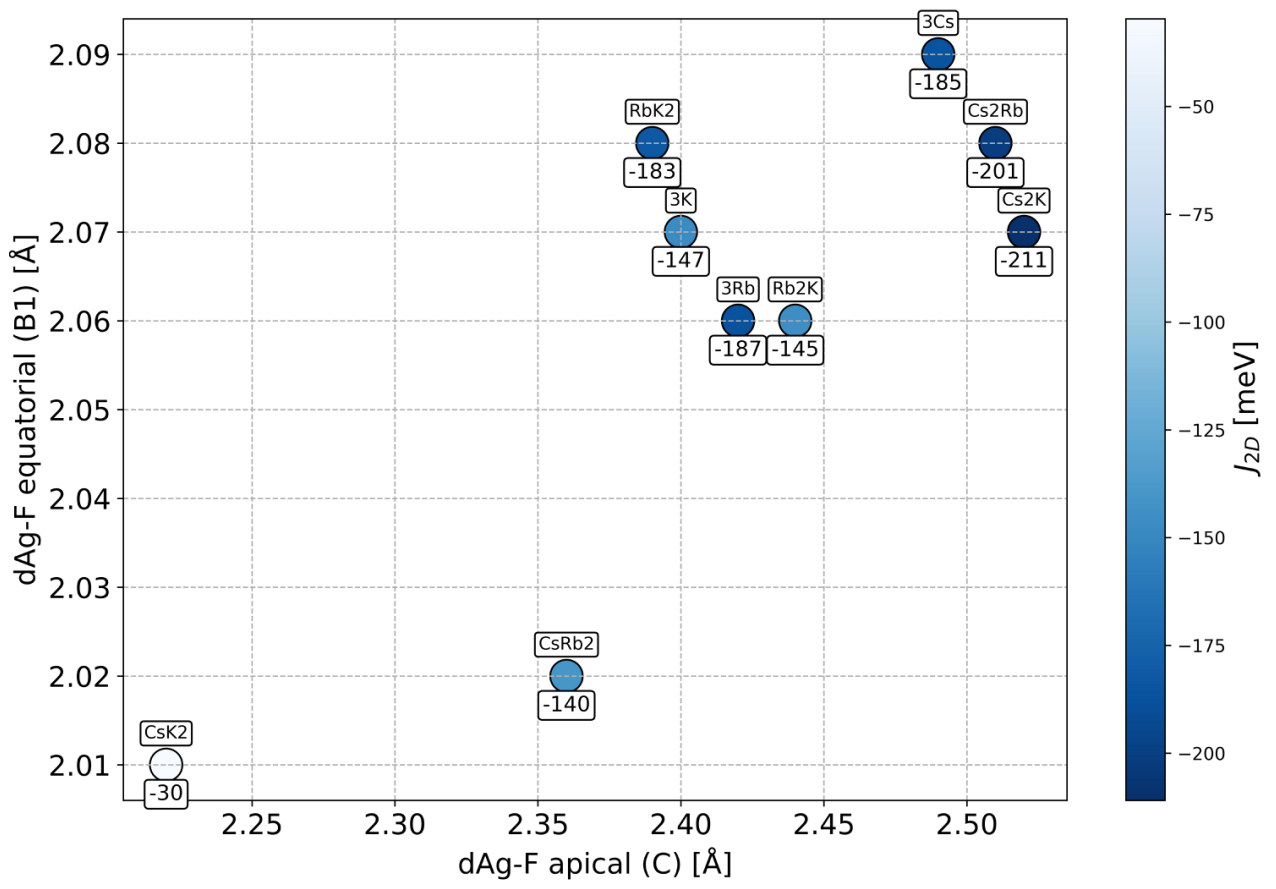

Figure 5. Dependence between Ag-F apical (**C**), Ag-F equatorial (**B1**) and $J_{2D}$ parameters for $[AgF_2]$ monolayer in orthorhombic *Pnma* phases. The labels at the points refer to the $J_{2D}$ values (lower) and the formula of the compound (upper). For CsK2 and CsRb2 phases (**B1**/**B2** <1) we considered shorter Ag-F equatorial bond lengths. Precise values are presented in Table SI1 in ESI.

Generally, it can be observed that the presence of the AFD in the $[AgF_2]$ layer (**B1/B2**<1 or $\psi \neq 0$) greatly reduces antiferromagnetic superexchange. The $J_{2D}$ value of –30 meV is calculated for the CsK2 system, where **B1/B2** equals 0.89. This is due to the unfavorable (antiferro) orbital ordering in the plane as shown in Ref.[21]. For the CsRb2 system, where the distortion is less pronounced compared to CsK2 (c.f. Fig. 4), the $J_{2D}$ (–140 meV) is nearly triple compared to the computed value for the bulk $AgF_2$ (see Table SI1, Figure SI1 in the ESI). Despite the distortion, $J_{2D}$ approaches the one for $La_2CuO_4$ (HTSC oxocuprate precursor)[27]. The compounds lacking a non-distortive arrangement of the $[AgF_2]$ layer show the strongest antiferromagnetic SE, with $J_{2D}$ up to –211 meV, as calculated for Cs2K; here, Ag-F apical bond length (**C**) is equal to 2.52 Å and equatorial (**B1**) one to 2.07 Å. The SE for this system may be compared to that of –265 meV, calculated by the same method for $AgF_2$ monolayer deposited on an $RbMgF_3$ substrate[21]. Additionally, the $J_{2D}$ of $[AgF_2]$ for Cs2K computed with SCAN method is equal –256 meV (Table SI1 in ESI). This method is more reliable for theoretical estimation of magnetic SE[19,39]. Our results suggest that materials with extremely strong 2D antiferromagnetic interaction can be obtained in the bulk form rather than as a fragile single layer (achievable only via advanced nanotechnology in ultra-high vacuum conditions).

It should be also emphasized, that the dependence between the Ag-F-Ag bond angle and the $J_{2D}$ values is also observed in the investigated series. This is particularly notable for the Rb3 and Rb2K systems, where the chemical environment of the $[AgF_2]$ layer is identical (i.e. [RbF] neighbors $[AgF_2]$). For very similar values of **C** and **B1**, the superexchange constant for the former system is –187 meV, whereas for the latter, it is –145 meV. This can be explained by the Goodenough-Kanamori-Anderson (GKA) rules[10–12], where bond angle (**α**) itself plays crucial role in the level of *pd* hybridization and, therefore, the strength of SE interaction. For Rb3, this angle is 163.6°, while for Rb2K, it is 159.2°. The variation in angles stems from a slight difference in the Ag-Ag distance at the level of 0.03 Å. The greater the value **A**, the flatter the [$AgF_2$] layer, and the more pronounced $|J_{2D}|$ linked with the antiferromagnetic ordering.

As mentioned previously, in systems with $P2_1/c$ symmetry, two distinct silver(II) difluoride sheets can be identified due to their different chemical environments. Consequently, two $J_{2D}$ constants for each system have been determined. In almost all considered monoclinic systems (except for mRbK2), the proximity to the [KF] sheet induces antiferrodistortion of the [$AgF_2$] layer, diminishing the $|J_{2D}|$ values (Figure 6).

This is particularly apparent for mCsK2, where for [$AgF_2$] layer I, proximate to the [KF] sheet, $J_{2D}$ is estimated at –44 meV. However, the superexchange (SE) constant for the layer adjacent to [CsF] is more than fourfold greater, –192 meV (layer II). The largest $|J_{2D}|$ value ($J_{2D}$ = –218meV) is predicted for [$AgF_2$] layer I, at the proximity to [CsF], for mIII_RbCsK system, where **C** = 2.57Å, **B1** = 2.05Å. These values are close to the parameters **C** and **B1** for orthorhombic Cs2K, where [$AgF_2$] also neighbors [CsF] sheet and a calculated $J_{2D}$ for the latter is –211 meV. However, the second silver difluoride layer in mIII_RbCsK system, exhibits smaller $|J_{2D}|$, with the $J_{2D}$ value of –183meV.

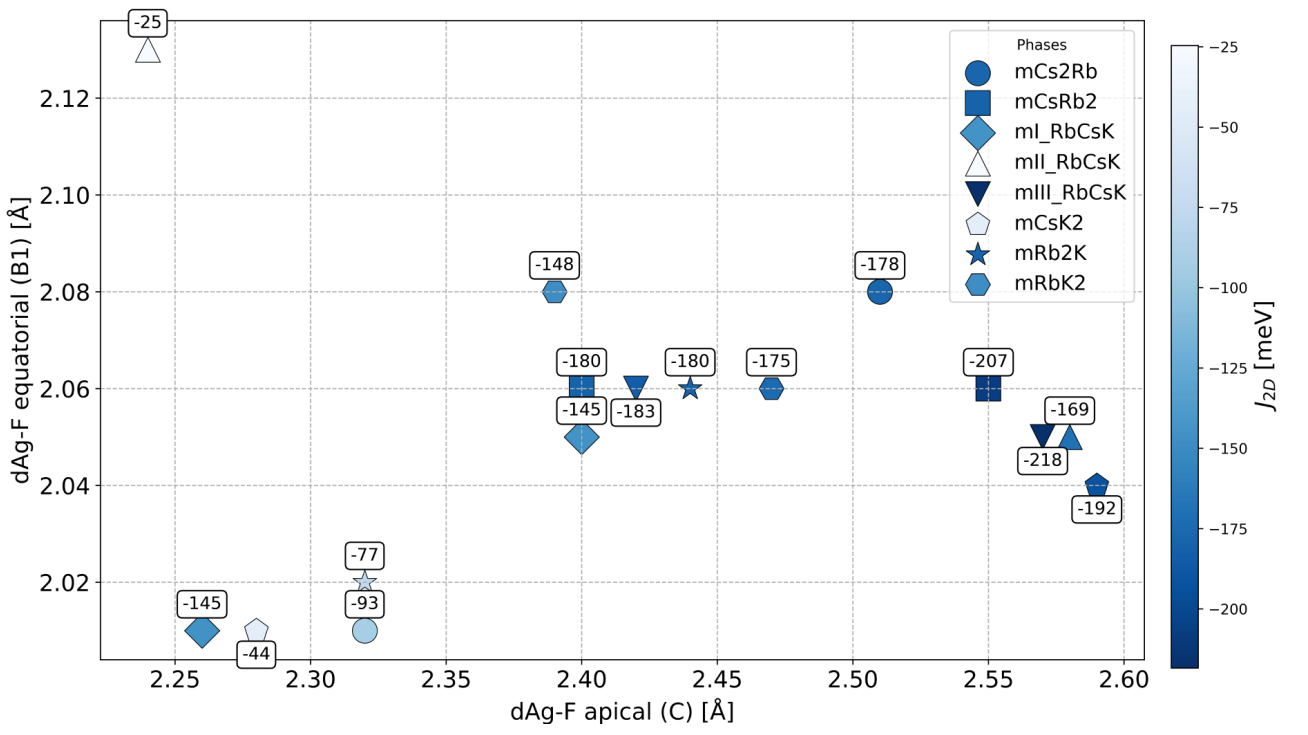

Figure 6. Dependence between apical Ag-F bond length (**C**), equatorial Ag-F bond length (**B1**) and $J_{2D}$ parameter for [$AgF_2$] monolayer in monclinic $P2_1/c$ phases. The numerical labels at each point refer to the $J_{2D}$ value. For phases with antiferrodistortive arrangement of [$AgF_2$] sheets and two equatorial bond lengths, (**B1/B2** <1), values of the shorter bond length (**B1**) were used.

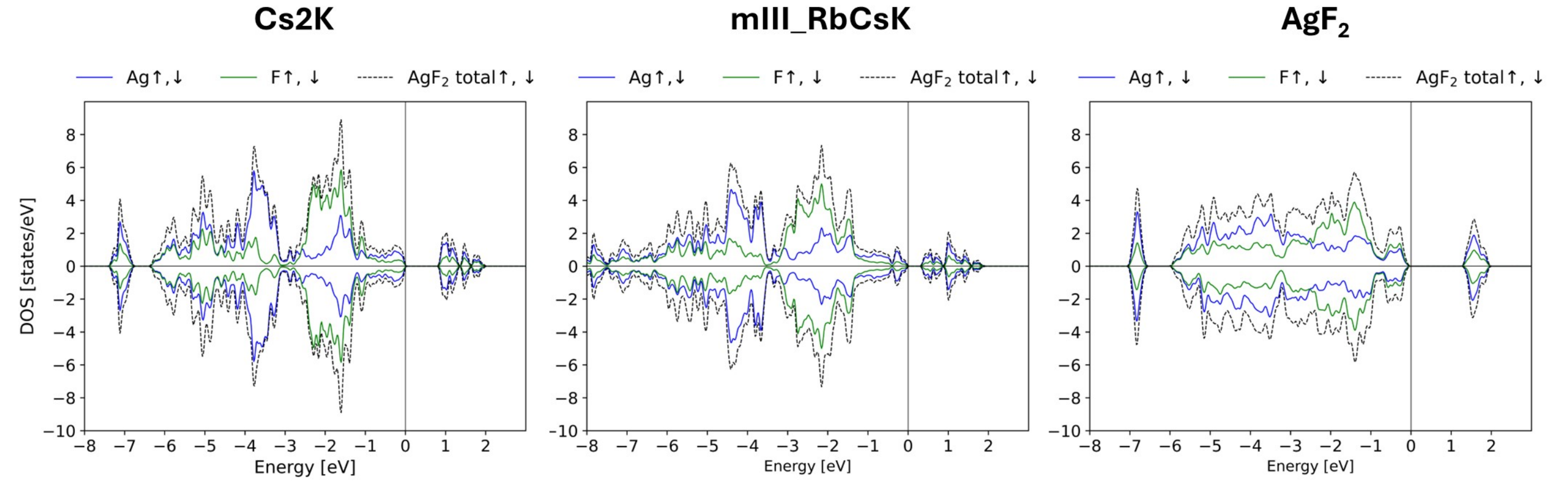

Figure 7. Electronic DOS for Cs2K, mIII_RbCsK and $AgF_2$, calculated using DFT+U method ($U_{Ag}$=5eV). In case of mIII_RbCsK, the Ag and F states from both [$AgF_2$] layers (I and II) are merged. Only Ag and F states from the [$AgF_2$] layers are shown for clarity.

In Figure 7 we present electronic density of states (DOS) for orthorhombic Cs2K and monoclinic mIII_RbCsK systems which showed the record large $|J_{2D}|$ values. Their DOS is compared to that for $AgF_2$ in the bulk form, calculated within DFT+U framework ($U_{Ag}$=5eV).

Table 3. Fundamental band gap (BG), the width of UHB, $J_{2D}$ values and magnetic moments (only for $Ag^{2+}$) for Cs2K, mIII_RbCsK and $AgF_2$. Results for DFT+U method ($U_{Ag}$=5eV). *for the [$AgF_2$] layers I / II, respectively.

| Parameter | mIII_RbCsK | Cs2K | $AgF_2$ |
|---|---|---|---|
| Magnetic mom. [$\mu_B$] | +/– 0.49 | +/– 0.48 | +/– 0.57 |
| Energy gap [eV] | 0.18 | 0.65 | 1.22 |
| UHB [eV] | 1.62 | 1.24 | 0.80 |
| $J_{2D}$ [meV] | –218/–183* | –211 | –52 |

Intergrowth structures show significant narrowing of the fundamental band gap (BG), as compared to bulk $AgF_2$. For Cs2K the BG is equal 0.65 eV – almost twice smaller than computed for bulk $AgF_2$ (1.22 eV, see Table 3). In the case of mIII_RbCsK, the computed BG is almost vanishing with the value of 0.18 eV. In terms of the Zaanen-Sawatzky-Allen (ZSA) classification[55,56], $AgF_2$ is a charge-transfer insulator, as documented by major contribution from the F states to the valence band. Interestingly, for the two intergrowth structures the predominant share to the valence band comes from silver, which place these systems near the boundary between charge-transfer and Mott-Hubbard insulators. For hypothetical structures, the level of $Ag_d$-$F_p$ hybridization is significant, as indicated by the large contribution from fluorine states to both the valence and the conduction band (or Upper Hubbard Band, UHB, Figure 5). Increased covalence is also behind the reduction of the value of magnetic moment of the $Ag^{2+}$ ion (Table 3) as compared to that in $AgF_2$ (0.57 μB)[35,36]; in mIII_RbCsK and Cs2K magnetic moment of $Ag^{2+}$ is equal 0.49 $\mu_B$ and 0.48 $\mu_B$, respectively pointing to even stronger *pd* mixing than for $AgF_2$. This is in line with good orbital overlap at the 180° bridge geometry, and it is obviously reflected in the large $|J_{2D}|$ values in both systems (Table 3).

The width of the (UHB) for both hypothetical flat-layer structures is much larger than for bulk $AgF_2$. Specifically, for mIII_RbCsK and Cs2K, it is equal 1.62 eV and 1.24 eV, respectively, whereas for silver(II) difluoride it is 0.80 eV. This is beneficial as it reduces tendency for polaron formation upon electron doping[21,37].

**D Energetic and dynamic stability of the intergrowth compounds**

We calculated the relative energetic stability of the $P2_1/c$ and *Pnma* phases with the same stoichiometries. Additionally, we calculated the energy effects associated with the formation of both types of intergrowth structures, considering four potential synthetic paths (Table 4, further details in ESI). In this manner, we determined the energies of formation for all hypothetical systems starting from four different sets of substrates (cf. reaction equations in section S.II in ESI), $dE_N$, where N=1-4.

A negative value of specific $dE_x$ indicates the energetic stability of the intergrowth compound over the substrates used for a given reaction. Interestingly, calculations reveal that all *Pnma* intergrowth systems are energetically favored over the substrates in case of reaction 4 (negative $dE_4$). Here, $M_2AgF_4$ (M=K, Rb, Cs), $MMgF_3$ (M=K, Rb, Cs) and $AgF_2$ are considered as substrates (see ESI). It suggests that, if auxiliary reactions will not be kinetically facile, starting from this specific set of compounds, one might possibly reach a given intergrowth system. Importantly, the Cs2K system with largest $|J_{2D}|$ is energetically preferred over substrates in the case of $dE_1$ ($2CsMgF_3$+$KAgF_3$) and $dE_4$ ($Cs_2AgF_4$+$KMgF_3$+$MgF_2$). Obviously, in more precise considerations, zero-point energy as well as entropy factor (potentially favoring substrate stability over products at higher temperatures) should be considered.

The obtained data suggest higher energetic stability for three monoclinic systems in comparison to their orthorhombic counterparts with the same stoichiometry. Specifically, for mRbK2, mCsRb2, and mCsK2, the energies were found to be lower by −169, −184, and −260 meV/f.u., respectively, than those of their orthorhombic forms. In general, systems (*Pnma* or $P2_1/c$) with large intra-sheet superexchange constants are energetically more favored. As an example, the monoclinic mCsRb2 with $J_{2D}$ of −180 meV and −207 meV, is favored by 183 meV over the *Pnma* variant – where $J_{2D}$ is equal −139 meV. Clearly, magnetic ordering strongly contributes to stability in these cases. This may be considered as a rare example where stability is dependent on the strength of magnetic interactions in different polymorphic forms.

Since the dynamic (lattice) stability is crucial for its successful preparation of any chemical system, we have decided to assess the dynamic stability and phonon dispersion for the extremely strongly coupled orthorhombic system Cs2K (Figure 8). The calculations suggest the absence of imaginary phonons throughout the entire dispersion range, including high symmetry points in the Brillouin zone (Figure 6, right). The resulting lattice stability at p → 0 and T → 0 renders this system a local minimum on the energy surface, which is a prerequisite for its possible synthesis.

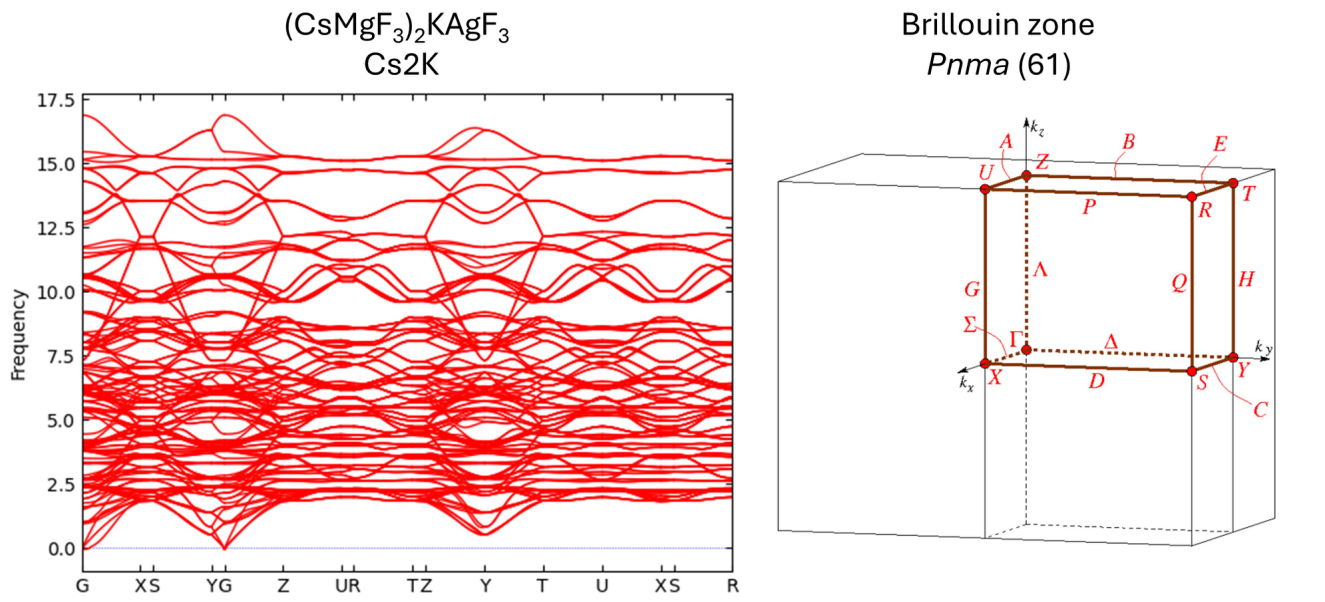

Figure 8. Phonon dispersion for the orthorhombic Cs2K (*Pnma*) system. On the right side of the figure, the Brillouin zone with marked points of high symmetry for the *Pnma* space group[57].

Table 4. Comparison of the formation energies, $dE_N$, where N=1-4, for all studied intergrowth systems. Entries for which specific reaction equations are not applicable for a given compound are highlighted as N.A. Relative stability is defined as $dE = E_{P21/c}-E_{Pnma}$ [meV/f.u.]. "SG" stands for Space Group.

| SG | Short name | Formula | Relative stability [meV] | $dE_1$ [meV/f.u.] | $dE_2$ [meV/f.u.] | $dE_3$ [meV/f.u.] | $dE_4$ [meV/f.u.] |
|---|---|---|---|---|---|---|---|
| *Pnma* | Cs3 | $(CsMgF_3)_2CsAgF_3$ | N.A. | +148 | N.A. | +243 | −177 |
| | Cs2Rb | $(CsMgF_3)_2RbAgF_3$ | N.A. | −217 | +191 | +286 | −135 |
| | CsRb2 | $(RbMgF_3)_2CsAgF_3$ | N.A. | +588 | +181 | −115 | −238 |
| | Cs2K | $(CsMgF_3)_2KAgF_3$ | N.A. | −344 | +239 | +334 | −86 |
| | CsK2 | $(KMgF_3)_2CsAgF_3$ | N.A. | +808 | +225 | −299 | −137 |
| | Rb3 | $(RbMgF_3)_2RbAgF_3$ | N.A. | +194 | N.A. | −102 | −226 |
| | Rb2K | $(RbMgF_3)_2KAgF_3$ | N.A. | −52 | +123 | −172 | −296 |
| | RbK2 | $(KMgF_3)_2RbAgF_3$ | N.A. | +474 | +299 | −225 | −63 |
| | K3 | $(KMgF_3)_2KAgF_3$ | N.A. | +204 | N.A. | −320 | −158 |
| $P2_1/c$ | mCs2Rb | $(CsMgF_3)_2RbAgF_3$ | +180 | −37 | +371 | +466 | +45 |
| | mCsRb2 | $(RbMgF_3)_2CsAgF_3$ | −184 | +405 | −3 | −299 | −422 |
| | mI_RbCsK | $RbCsKMg_2AgF_9$ | N.A. | +506 | +330 | +913 | N.A. |
| | mII_RbCsK | $RbCsKMg_2AgF_9$ | N.A. | +103 | −72 | +511 | N.A. |
| | mIII_RbCsK | $RbCsKMg_2AgF_9$ | N.A. | +37 | −139 | +445 | N.A. |
| | mCsK2 | $(KMgF_3)_2CsAgF_3$ | −260 | +548 | −35 | −559 | −397 |
| | mRb2K | $(RbMgF_3)_2KAgF_3$ | +167 | +115 | +291 | −5 | −128 |
| | mRbK2 | $(KMgF_3)_2RbAgF_3$ | −169 | +306 | +130 | −394 | −232 |

## E Doping approaches and estimate of resulting $T_c$ values

A natural question arising from our results is what is the prospect of hypothetical systems studied here for superconductivity i.e. what would be their $T_C$ values if optimally doped. To estimate the expected $T_C$ values for intergrowth fluoride structures we have adopted here the empirical relationship between $T_c$ and $J_{2D}$ for non-interacting layers in cuprates – precursors of the HTSC[9]. A similar relation was previously applied to single $[AgF_2]$ layers placed on appropriate fluoride structures[21].

The application of the $T_c$ on $J_{2D}$ dependence (Figure 9) indicates that $[AgF_2]$ layers may exhibit even higher $T_c$ values, than those known for single-layer oxocuprate HTSCs. E.g., the optimally doped mIII_RbCsK system (layer I) can reach a $T_c$ of 148 K, based on DFT+U calculations. Considering the $J_{2D}$ from SCAN method, for Cs2K system (*cf.* TABLE SI1 in the ESI), a $T_c$ of approximately 200 K is anticipated at optimum doping.

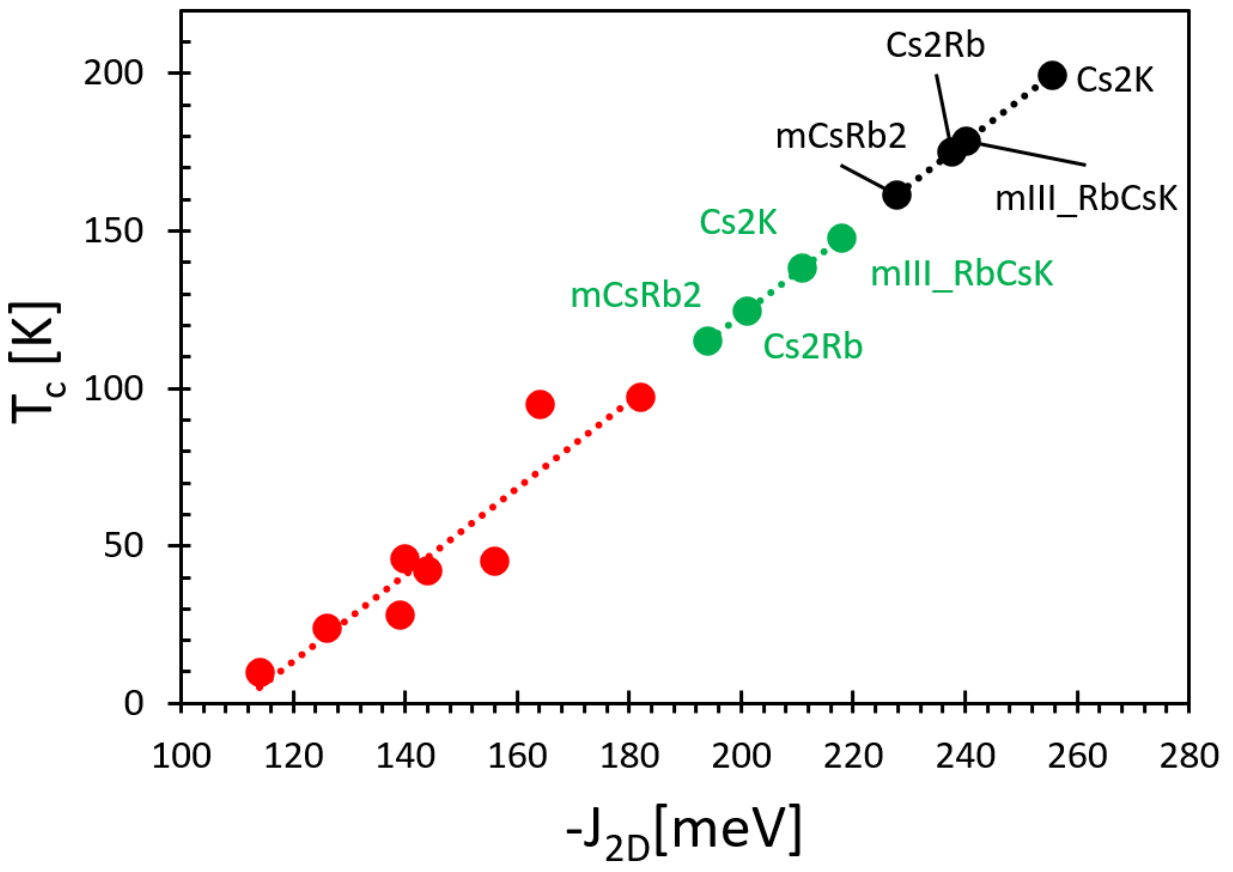

Figure 9. $T_c$ as a function of $J_{2D}$ for cuprates (red dots)[9] with extrapolation to hypothetical intergrowth compounds, where green dots are for DFT+U and black – for SCAN results of $J_{2D}$.

How could one dope such bulk intergrowth systems? To illustrate this let us focus on the most promising candidate as a HTSC precursor, namely orthorhombic Cs2K system. Doping of the $[AgF_2]$ layers could be accomplished through the following substitutions: $K^+ \rightarrow Ba^{2+}$ in [MF] charge-reservoir layer (metal Y in Figure 1) or substituting $Mg^{2+}$ by $Li^+$ in the $[MgF_2]$ layer. The first approach is intended to inject electrons to the conduction band, while the latter to dope holes to the valence band. The choice of cation pairs for such substitutions stems from similarities of the ionic radii of $Ba^{2+}$ and $K^+$ as well as $Mg^{2+}$ and $Li^{+}$[58].

In the case of electron doping, through gradual substitution of $K^+$ by $Ba^{2+}$, several levels of doping (x) were considered, with x = 1, 2.5 and from 5% to 25% with the increment of 5%. Unit cells were fully optimized to detect the structural effects of such doping. The crucial aspect concerns the possible introduction of electrons into the $d_{(x2-y2)}$ states of $Ag^{2+}$ (UHB)[21,40,41]. Therefore, DOS calculations were performed, and the results are presented in the Figure 10.

As shown in Figure 10, with the gradual substitution of $K^+$ by $Ba^{2+}$, the electron density emerges at the Fermi level (set to zero) within the UHB band. To evaluate the doping level ($\delta n_{Ag(UHB)}$), specifically the density of the occupied states within this band, we calculated the ratio of the integrated populated part to the total population of UHB (further details can be found in part IV in ESI). For the investigated range of cation substitution, the $\delta n_{Ag(UHB)}$ (doping level) of up to 23% is achieved (see Figure 11, A and Table SI3 in ESI). The divergence between the theoretical value and the one obtained from the integration (for example, 25% vs. 23%) reflects the typical error of the latter method.

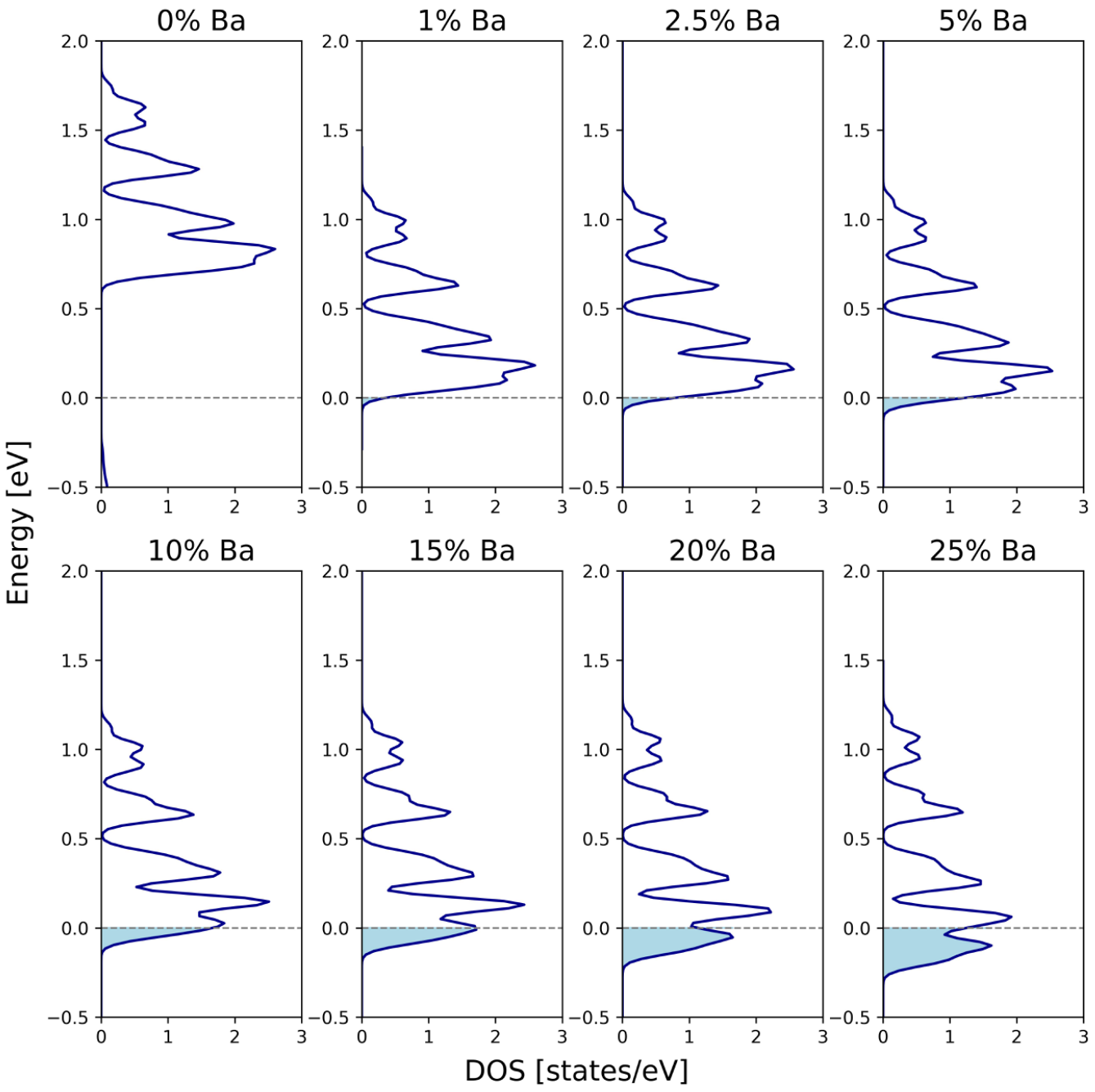

Figure 10. Density of states for $d_{(x2-y2)}$ orbital (UHB) of silver from $[AgF_2]$ layer. The Upper left for the pristine Cs2K compound, lower right for 25% of $Ba^{2+}$ substitution of $K^+$ in Cs2K structure. Fermi level denoted as 0.

To place these findings in broader context, the optimal electron doping level for bulk $AgF_2$ was estimated to be 5%[59]. However, it should be noted, that bulk $AgF_2$ features strongly puckered $[AgF_2]$ layers – structurally quite different from the flat layers in Cs2K. This implies that the optimal doping level for achieving high-temperature superconductivity may actually vary in case of that intergrowth compound[37,40]. The observed changes in the population level of UHB band, are also reflected in the changes in magnetic moment of $Ag^{2+}$ in the layer. For the pristine Cs2K compound, silver cation exhibits moment of $0.48\mu_B$; with 5% Ba substitution, of $0.46\ \mu_B$; and of $0.29\ \mu_B$ for 25% Ba concentration level (Figure 11, D). It is similar to the computed reduction of $Ag^{2+}$ magnetic moments in $[AgF_2]$ monolayer with electron doping in chemical capacitor setup[40,41]. The calculations also reveal another important aspect related to doping. It can be observed, that with $Ba^{2+}$ doping, the UHB widens slightly comparing to the undoped compound. At the same time, there are no noticeable changes in the F-Ag-F bond angle (**α**) or in the Ag-Ag distances (**A**). This constitutes a difference between the bulk system studied here and a $[AgF_2]$ monolayer in chemical capacitor setup[40], where **α** values are smaller for higher degrees of doping.

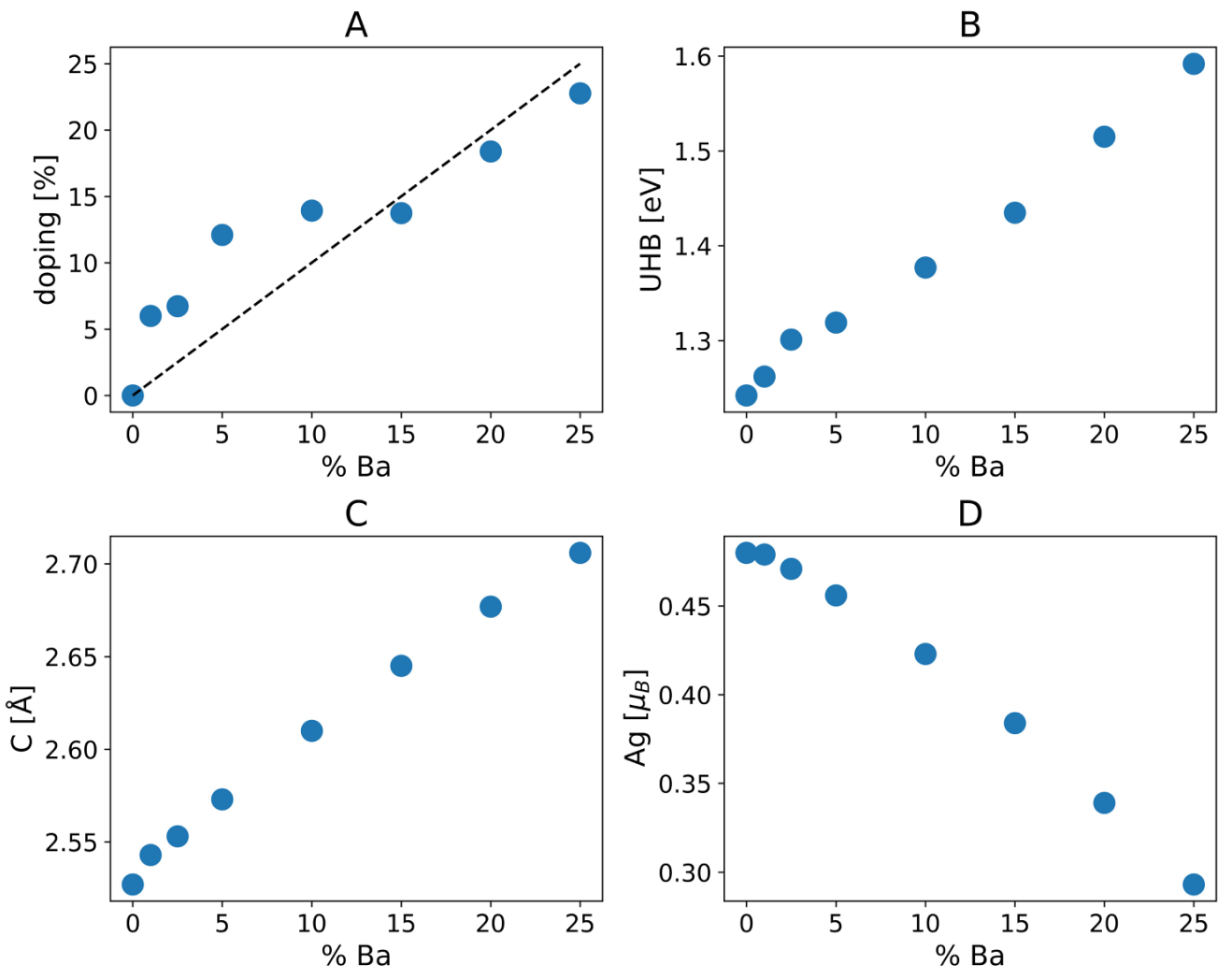

Figure 11. Dependence of A) doping level calculated from integration of the UHB, $\delta n_{Ag(UHB)}$, vs. % Ba, where dotted line shows expected linear dependence B) UHB width vs. % Ba, C) apical Ag-F bond length (C) vs. % Ba and D) $Ag^{2+}$ magnetic moment vs. % Ba for Cs2K. Numerical values are shown in the Table SI3 in ESI.

Changes of cell parameters upon doping are also seen. In the intergrowth structure, the largest unit cell vector *b* enlarges, and the apical Ag-F bond **C** noticeably lengthens upon $K^+ \rightarrow Ba^{2+}$ doping (Figure 11, and Table SI3 in ESI).

In the case of hole doping ($h^+$), we considered the gradual substitution of $Mg^{2+}$ by $Li^+$ within the [$MgF_2$] layer, where x = 1, 2.5, and from 5 to 25% with increments of 5. Our attention was focused on the depopulation of silver $d_{(x^2-y^2)}$ β-states, which is crucial for achieving HTSC. However, as shown in Figure 9, depopulation of these states across the entire range of $Li^+$ concentration is not the only effect seen; depopulation of Ag$d_{(z^2)}$ and apical F$p_{(z)}$ orbitals, which are axially oriented perpendicular to the [$AgF_{4/2}$] sheets, is substantial. This is similar to what was observed in our previous study on possible hole doping to a flat $AgF_2$ single layer[40]. When the depopulation of Ag$d_{(z^2)}$ states is too large, it is undesirable for superconductivity, as it indicates a propensity towards stabilization of high-spin $Ag^{3+}$(S=1) upon doping. [60,61] Indeed, our VCA calculations with $La^{3+} \rightarrow Ba^{2+}$ partial substitution in $La_2CuO_4$ (section V of the ESI) suggest that for a 15% doping, $\delta n_{Cu(dz^2)}$ is 6.7% and $\delta n_O$ is 8.4% (apical oxygen).

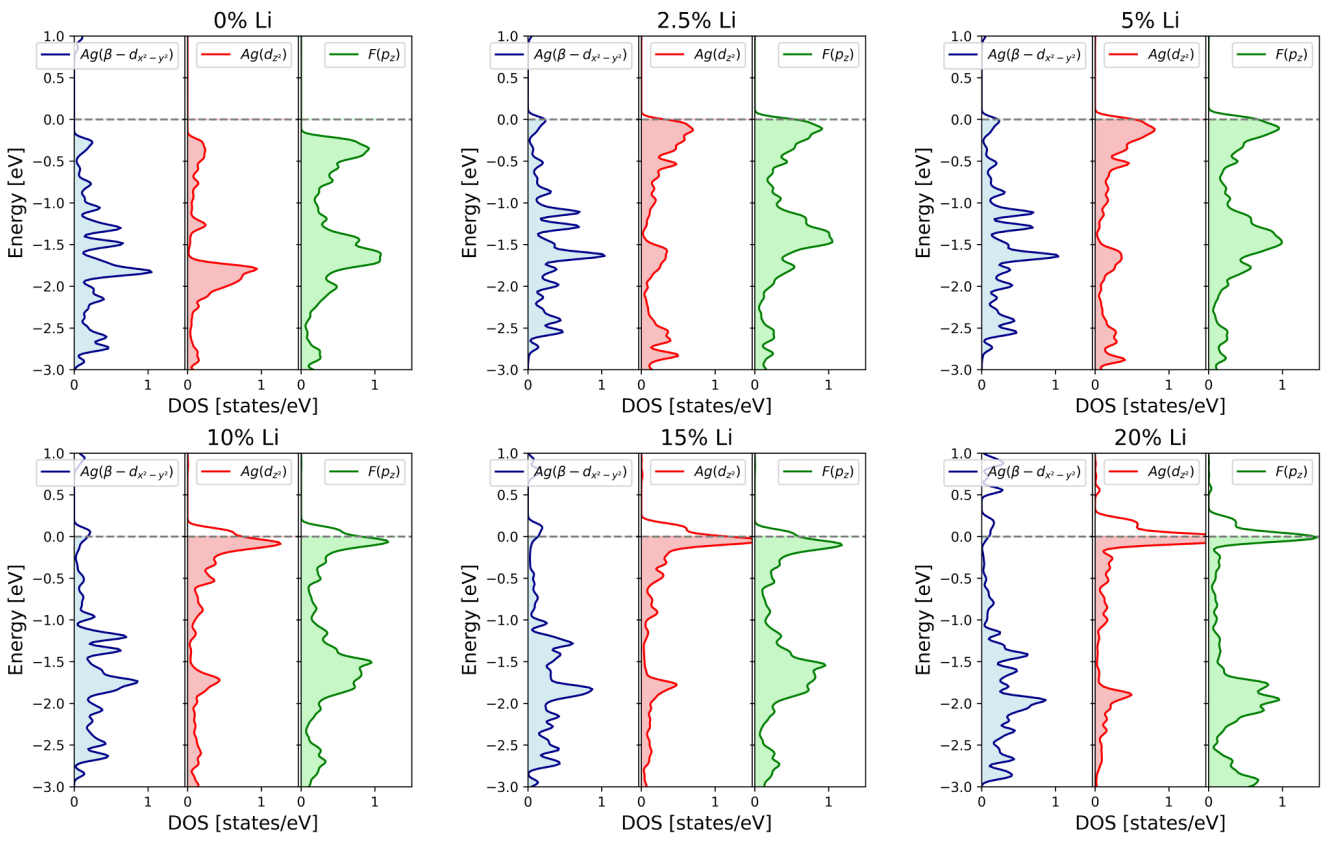

Figure 12. Density of states for $d_{(x2-y2)}$ (β state) (blue) and $d_{(z2)}$ orbitals of silver (red) and $p_{(z)}$ (green) orbital of apical fluorine in [$AgF_{4/2}$] sheet. The upper left for the pristine Cs2K compound, the lower right for 20% of $Li^+$ substitution of $Mg^{2+}$ in Cs2K structure. Fermi level set to 0 eV.

We estimated that $\delta n_{Ag(dz^2)}$ is 11.0% at a 20% concentration of Li, where simultaneously, β-$d_{(x^2-y^2)}$ ($\delta n_{Ag(\beta-d(x^2-y^2))}$, for beta states) equals 1.3%. For the F$p_{(z)}$ states, the depopulation level is 6.6% (for data at all Li concentrations see Table ESI4, Figure 12). Here, again, sum of values obtained from band integration (18.9%) diverges from theoretical value of 20% due to numerical errors in integration. Therefore, a decrease of Ag magnetic moment and a concomitant increase of F moment are expected – as it is indeed observed in this case. The modest doping levels of Ag and F are consistent with the relatively small changes in their magnetic moments. The magnetic moment of silver diminishes from 0.48 $\mu_B$ to 0.44 $\mu_B$, while for fluorine, it increases from 0.00 $\mu_B$ to 0.12 $\mu_B$ (Figure 13 and Table SI4), with increasing of Li concentration from 0% to 20% in Cs2K.

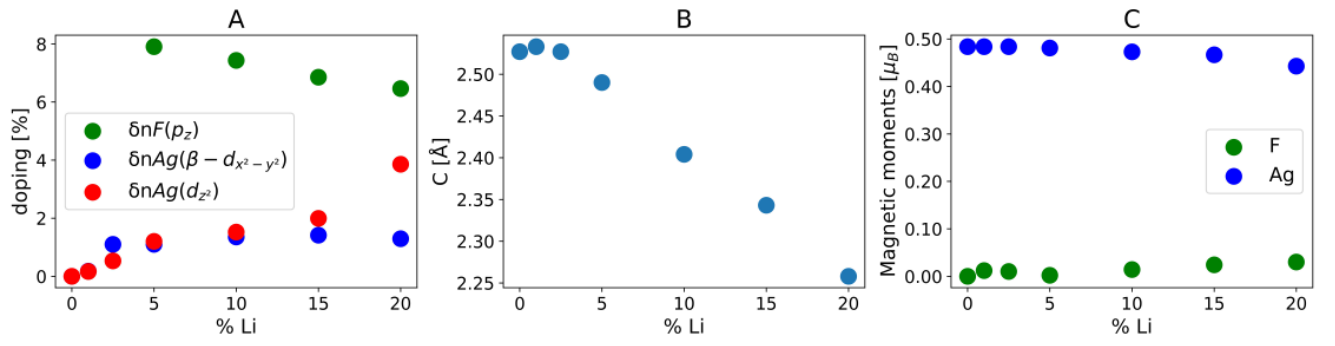

Figure 13. Dependence of A) doping level, $\delta n_{Ag(\beta-d(x^2-y^2))}$, $\delta n_{F(p_z)}$, $\delta n_{Ag(d_{z^2})}$ vs. % Li, B) apical Ag-F bond length (**C**) vs. % Li and C) Ag and F magnetic moment vs. doping level, %Li, for Cs2K. All values presented in the Table SI4 in ESI.

With gradual substitution of $Mg^{2+}$ by $Li^+$, the cell volume also decreases - a significant reduction of the *b* parameter is seen (Table SI4). Along with this, the apical Ag–F distance decreases from 2.53 Å to 2.26 Å, from x=0% to 20% of Li.

At 25% Li substitution, the geometry of the [$AgF_6$] octahedra changes considerably; the first coordination sphere of silver is now a compressed octahedron (as in inversed Jahn-Teller effect) with short apical Ag-F bonds and longer equatorial ones (**C** = 2.07 Å while **B1=B2** = 2.09 Å, see Table SI4). Consequently, the $d_{(z^2)}$ states are now upper in energy than the $d_{(x^2-y^2)}$ states of silver, which is the opposite to the case of elongated octahedra (normal Jahn-Teller effect). Therefore, the major share of holes is located in the $d_{(z^2)}$ orbital of Ag, while the $d_{(x^2-y^2)}$ are only slightly depopulated (see Figure SI10 in ESI). This means that achieving the desired hole-doping level of 14%[40] to Ag$d_{(x^2-y^2)}$ states in these hypothetical intergrowth systems via $Mg^{2+} \rightarrow Li^+$ substitution will not be easy, while electron doping connected with $K^+ \rightarrow Ba^{2+}$ substitution should be more facile.

## Conclusions

We have studied theoretically a handful of several new hypothetical intergrowth fluorides featuring [$AgF_2$] layers. Depending on the stoichiometry of the intergrowth cells, the [$AgF_2$] layers exhibit different structural properties. For several stoichiometries nearly flat layers are stabilized, free from antiferrodistortive bond pattern. Landau analysis of structural phase transitions has allowed us to draw a generalized phase diagram of the [$AgF_2$] sheet subject to chemical strain. The phase diagram shows sheet buckling for positive strain as well as the undesired AFD bond pattern (antiferro orbital ordering) for negative strain. Fortunately, an ideal window exists for intermediate strain where ideal flat sheets with ferro orbital ordering are possible.
Such layers display magnetic and electronic properties similar to those of precursors of oxocuprate HTSCs. Based on the calculated intra-sheet magnetic SE values exceeding –250 meV (SCAN) for the best structures the estimated critical superconducting temperatures ($T_c$) up to 200 K surpass those achieved for cuprates. Theoretical study of partial $K^+ \rightarrow Ba^{2+}$ substitution in Cs2K system indicates a possibility for genuine electron doping of $Ag^{2+}$ in [$AgF_2$] layer. However, the hole doping of planar orbitals in the [$AgF_2$] monolayer seems to be difficult to achieve. The substitution of $Mg^{2+}$ with $Li^+$ primarily leads to the depopulation of the F$p_{(z)}$ and Ag$d_{(z2)}$ states; the degree of hole doping in the $d_{(x2-y2)}$ is small.

The Cs2K system with the record large $|J_{2D}|$ value shows dynamic (phonon) stability and thus it is a local minimum on the potential energy surface. Several different reaction pathways towards Cs2K system were studied and some are downhill in energy, highlighting the possibility of synthesis of this and related intergrowth compounds.

## Author contributions

W.G. and J.L. planned the research and provided funding. D.J. performed calculations, analysed data, and wrote the initial version of the manuscript. W.G. and J.L. performed additional analyses and shaped the final version of the manuscript.

## Conflicts of interest

There are no conflicts to declare.

## Acknowledgements

W.G. is grateful to the Polish National Science Center (NCN) for project Maestro (2017/26/A/ST5/00570). Quantum mechanical calculations were performed using supercomputing resources of the ICM UW (project SAPPHIRE [GA83-34]). J.L. acknowledges support from MIUR Italian Ministry for Research through PRIN Project No. 20207ZXT4Z.

## Bibliography

1 A. Schilling, M. Cantoni, J. D. Guo and H. R. Ott, *Superconductivity above 130 K in the Hg–Ba–Ca–Cu–O system | Nature, Nature.*
2 J. G. Bednorz and K. A. Müller, *Possible highTc superconductivity in the Ba–La–Cu–O system, Zeitschrift für Physik B Condensed Matter, 1986,* ***64****, 189–193.*
3 G. Wu, Y. L. Xie, H. Chen, M. Zhong, R. H. Liu, B. C. Shi, Q. J. Li, X. F. Wang, T. Wu, Y. J. Yan, J. J. Ying and X. H. Chen, *Superconductivity at 56 K in samarium-doped SrFeAsF, Journal of Physics: Condensed Matter, 2009,* ***21****, 142203.*
4 B. Cheng, D. Cheng, K. Lee, L. Luo, Z. Chen, Y. Lee, B. Y. Wang, M. Mootz, I. E. Perakis, Z.-X. Shen, H. Y. Hwang and J. Wang, *Evidence for d-wave superconductivity of infinite-layer nickelates from low-energy electrodynamics, Nature Materials, 2024, 1–7.*
5 J. Cheng, J. Bai, B. Ruan, P. Liu, Y. Huang, Q. Dong, Y. Huang, Y. Sun, C. Li, L. Zhang, Q. Liu, W. Zhu, Z. Ren and G. Chen, *Superconductivity in a Layered Cobalt Oxychalcogenide Na2CoSe2O with a Triangular Lattice, Journal of the American Chemical Society, 2024,* ***146****, 5908–5915.*
6 J.-F. Ge, Z.-L. Liu, C. Liu, C.-L. Gao, D. Qian, Q.-K. Xue, Y. Liu and J.-F. Jia, *Superconductivity above 100 K in single-layer FeSe films on doped SrTiO3, Nature Materials, 2015,* ***14****, 285–289.*
7 Y. J. Song, J. B. Hong, B. H. Min, Y. S. Kwon, K. J. Lee, M. H. Jung and J. S. Rhyee, *Superconducting properties of a stoichiometric FeSe compound and two anomalous features in the normal state, Journal of the Korean Physical Society, 2011,* ***59****, 312–316.*
8 F. C. Zhang and T. M. Rice, *Effective Hamiltonian for the superconducting Cu oxides, Physical Review B, 1988,* ***37****, 3759–3761.*
9 I. de P. R. Moreira, D. Muñoz, F. Illas, C. de Graaf and M. A. Garcia-Bach, *A relationship between electronic structure effective parameters and Tc in monolayered cuprate superconductors, Chemical Physics Letters, 2001,* ***345****, 183–188.*
10 J. B. Goodenough, *Theory of the Role of Covalence in the Perovskite-Type Manganites $LaMnO_3$, Physical Review, 1955,* ***100****, 564–573.*
11 P. W. Anderson, *Antiferromagnetism. Theory of Superexchange Interaction, Physical Review, 1950,* ***79****, 350–356.*
12 J. Kanamori, *Superexchange interaction and symmetry properties of electron orbitals, Journal of Physics and Chemistry of Solids, 1959,* ***10****, 87–98.*
13 W. D. Wise, M. C. Boyer, K. Chatterjee, T. Kondo, T. Takeuchi, H. Ikuta, Y. Wang and E. W. Hudson, *Charge-density-wave origin of cuprate checkerboard visualized by scanning tunnelling microscopy, Nature Physics, 2008,* ***4****, 696–699.*
14 L. Braicovich, L. J. P. Ament, V. Bisogni, F. Forte, C. Aruta, G. Balestrino, N. B. Brookes, G. M. De Luca, P. G. Medaglia, F. M. Granozio, M. Radovic, M. Salluzzo, J. van den Brink and G. Ghiringhelli, *Dispersion of Magnetic Excitations in the Cuprate $La_2CuO_4$ and $CaCuO_2$ Compounds Measured Using Resonant X-Ray Scattering, Physical Review Letters, 2009,* ***102****, 167401.*
15 A. S. Botana and M. R. Norman, *Layered palladates and their relation to nickelates and cuprates, Physical Review Materials, 2018,* ***2****, 104803.*
16 B. Lilia, R. Hennig, P. Hirschfeld, G. Profeta, A. Sanna, E. Zurek, W. E. Pickett, M. Amsler, R. Dias, M. I. Eremets, C. Heil, R. J. Hemley, H. Liu, Y. Ma, C. Pierleoni, A. N. Kolmogorov, N. Rybin, D. Novoselov, V. Anisimov, A. R. Oganov, C. J. Pickard, T. Bi, R. Arita, I. Errea, C. Pellegrini, R. Requist, E. K. U. Gross, E. R. Margine, S. R. Xie, Y. Quan, A. Hire, L. Fanfarillo, G. R. Stewart, J. J. Hamlin, V. Stanev, R. S. Gonnelli, E. Piatti, D. Romanin, D. Daghero and R. Valenti, *The 2021 room-temperature superconductivity roadmap, J. Phys.: Condens. Matter, 2022,* ***34****, 183002.*
17 M. R. Norman, *Materials design for new superconductors, Reports on Progress in Physics, 2016,* ***79****, 074502.*
18 W. Grochala and R. Hoffmann, *Real and Hypothetical Intermediate-Valence AgII/AgIII and AgII/AgI Fluoride Systems*

*as Potential Superconductors, Angewandte Chemie International Edition, 2001,* ***40****, 2742–2781.*
*19 C. Miller and A. S. Botana, Cupratelike electronic and magnetic properties of layered transition-metal difluorides from first-principles calculations, Physical Review B, 2020,* ***101****, 195116.*
*20 W. Grochala, Silverland: the Realm of Compounds of Divalent Silver—and Why They are Interesting, Journal of Superconductivity and Novel Magnetism, 2018,* ***31****, 737–752.*
*21 A. Grzelak, H. Su, X. Yang, D. Kurzydłowski, J. Lorenzana and W. Grochala, Epitaxial engineering of flat silver fluoride cuprate analogs, Physical Review Materials, 2020,* ***4****, 084405.*
*22 R. Ofer, G. Bazalitsky, A. Kanigel, A. Keren, A. Auerbach, J. S. Lord and A. Amato, Magnetic analog of the isotope effect in cuprates, Phys. Rev. B, 2006,* ***74****, 220508.*
*23 D. I. Khomskii, Basic Aspects of the Quantum Theory of Solids: Order and Elementary Excitations, Cambridge University Press, Cambridge, 2010.*
*24 D. Kurzydłowski, T. Jaroń, A. Ozarowski, S. Hill, Z. Jagličić, Y. Filinchuk, Z. Mazej and W. Grochala, Local and Cooperative Jahn–Teller Effect and Resultant Magnetic Properties of M2AgF4 (M = Na–Cs) Phases, Inorganic Chemistry, 2016,* ***55****, 11479–11489.*
*25 R. -H. Odenthal and R. Hoppe, Uber Fluoroargentate(II) $M^{II}$ $[AgF_4]$ mit $M^{II}$ = Ba, Sr, Ca, Hg, Cd, sowie $Ba_2AgF_6$, Zeitschrift anorg allge chemie, 1971,* ***385****, 92–101.*
*26 S. E. McLain, M. R. Dolgos, D. A. Tennant, J. F. C. Turner, T. Barnes, T. Proffen, B. C. Sales and R. I. Bewley, Magnetic behaviour of layered Ag(II) fluorides, Nature Materials, 2006,* ***5****, 561–565.*
*27 D. Kurzydłowski and W. Grochala, Prediction of Extremely Strong Antiferromagnetic Superexchange in Silver(II) Fluorides: Challenging the Oxocuprates(II), Angewandte Chemie, 2017,* ***129****, 10248–10251.*
*28 D. Kurzydłowski, Z. Mazej, Z. Jagličić, Y. Filinchuk and W. Grochala, Structural transition and unusually strong antiferromagnetic superexchange coupling in perovskite KAgF3, Chemical Communications, 2013,* ***49****, 6262.*
*29 Z. Mazej, E. Goreshnik, Z. Jagličić, B. Gaweł, W. Łasocha, D. Grzybowska, T. Jaroń, D. Kurzydłowski, P. Malinowski, W. Koźminski, J. Szydłowska, P. Leszczyński and W. Grochala, KAgF3, K2AgF4 and K3Ag2F7: important steps towards a layered antiferromagnetic fluoroargentate(II), CrystEngComm, 2009,* ***11****, 1702–1710.*
*30 R. -H. Odenthal, D. Paus and R. Hoppe, Zur Magnetochemie der Fluoroargentate(II): Messungen an $Ba[AgF_4]$, $Sr[AgF_4]$, $Ba_2AgF_6$ sowie $K[AgF_3]$ und $Cs[AgF_3]$, Zeitschrift anorg allge chemie, 1974,* ***407****, 151–156.*
*31 P. Charpin, P. Plurien and P. Mériel, Étude par diffraction de neutrons de la structure cristalline et magnétique de $AgF_2$, Bulletin de Minéralogie, 1970,* ***93****, 7–13.*
*32 J. Gawraczyński, D. Kurzydłowski, R. A. Ewings, S. Bandaru, W. Gadomski, Z. Mazej, G. Ruani, I. Bergenti, T. Jaroń, A. Ozarowski, S. Hill, P. J. Leszczyński, K. Tokár, M. Derzsi, P. Barone, K. Wohlfeld, J. Lorenzana and W. Grochala, Silver route to cuprate analogs, Proceedings of the National Academy of Sciences, 2019,* ***116****, 1495–1500.*
*33 K. B. Lyons, P. A. Fleury, J. P. Remeika, A. S. Cooper and T. J. Negran, Dynamics of spin fluctuations in lanthanum cuprate, Physical Review B, 1988,* ***37****, 2353–2356.*
*34 N. Bachar, K. Koteras, J. Gawraczyński, W. Trzciński, J. Paszula, R. Piombo, P. Barone, Z. Mazej, G. Ghiringhelli, A. Nag, K.-J. Zhou, J. Lorenzana, D. Van Der Marel and W. Grochala, Charge-Transfer and d d excitations in AgF 2, Physical Review Research, 2022,* ***4****, 023108.*
*35 R. Piombo, D. Jezierski, H. P. Martins, T. Jaroń, M. N. Gastiasoro, P. Barone, K. Tokár, P. Piekarz, M. Derzsi, Z. Mazej, M. Abbate, W. Grochala and J. Lorenzana, Strength of correlations in a silver-based cuprate analog, Physical Review B, 2022,* ***106****, 035142.*
*36 W. Grochala, R. G. Egdell, P. P. Edwards, Z. Mazej and B. Žemva, On the Covalency of Silver-Fluorine Bonds in Compounds of Silver(I), Silver(II) and Silver(III), ChemPhysChem, 2003,* ***4****, 997–1001.*
*37 S. Bandaru, M. Derzsi, A. Grzelak, J. Lorenzana and W. Grochala, Fate of doped carriers in silver fluoride cuprate analogs, Physical Review Materials, 2021,* ***5****, 064801.*
*38 A. Grzelak, M. Derzsi and W. Grochala, Defect Trapping and Phase Separation in Chemically Doped Bulk AgF $_2$, Inorganic Chemistry, 2021,* ***60****, 1561–1570.*
*39 D. Jezierski and W. Grochala, Polymorphism of two-dimensional antiferromagnets, $AgF_2$ and $CuF_2$, Physical Review Materials, 2024,* ***8****, 034407.*
*40 D. Jezierski, A. Grzelak, X. Liu, S. K. Pandey, M. N. Gastiasoro, J. Lorenzana, J. Feng and W. Grochala, Charge doping to flat AgF $_2$ monolayers in a chemical capacitor setup, Physical Chemistry Chemical Physics, 2022,* ***24****, 15705–15717.*
*41 A. Grzelak, J. Lorenzana and W. Grochala, Separation-Controlled Redox Reactions, Angewandte Chemie International Edition, 2021,* ***60****, 13892–13895.*
*42 S. Iijima, T. Ichihashi and Y. Kubo, Structural Variations on Superconductor Tl-Ba-Ca-Cu-O Oxides, Japanese Journal of Applied Physics, 1988,* ***27****, L817.*
*43 T. Hopfinger, O. O. Shcherban, P. Galez, R. E. Gladyshevkii, M. Lomello-Tafin, J. L. Jorda and M. Couach, Intergrowth of structures in the Tl–Ba–Ca–Cu–O system, Journal of Alloys and Compounds, 2002,* ***333****, 237–248.*
*44 G. Kresse and J. Furthmüller, Efficient iterative schemes for ab initio total-energy calculations using a plane-wave basis set, Physical Review B, 1996,* ***54****, 11169–11186.*
*45 J. P. Perdew, A. Ruzsinszky, G. I. Csonka, O. A. Vydrov, G. E. Scuseria, L. A. Constantin, X. Zhou and K. Burke, Restoring the Density-Gradient Expansion for Exchange in Solids and Surfaces, Physical Review Letters, 2008,* ***100****, 136406.*
*46 P. E. Blöchl, Projector augmented-wave method, Physical Review B, 1994,* ***50****, 17953–17979.*
*47 G. Kresse and D. Joubert, From ultrasoft pseudopotentials to the projector augmented-wave method, Physical Review B, 1999,* ***59****, 1758–1775.*
*48 A. I. Liechtenstein, V. I. Anisimov and J. Zaanen, Density-functional theory and strong interactions: Orbital ordering in Mott-Hubbard insulators, Physical Review B, 1995,* ***52****, R5467–R5470.*

*49 D. Jezierski, K. Koteras, M. Domański, P. Połczyński, Z. Mazej, J. Lorenzana and W. Grochala, Unexpected Coexisting Solid Solutions in the Quasi-Binary Ag(II)F2/Cu(II)F2 Phase Diagram, Chemistry – A European Journal, 2023,* ***29****, e202301092.*
*50 D. Jezierski, Z. Mazej and W. Grochala, Novel Ternary AgIICoIIIF5 Fluoride: Synthesis, Structure and Magnetic Characteristics, w druku, Dalton Transactions.*
*51 J. Heyd, G. E. Scuseria and M. Ernzerhof, Hybrid functionals based on a screened Coulomb potential, The Journal of Chemical Physics, 2003,* ***118****, 8207–8215.*
*52 J. Sun, A. Ruzsinszky and J. P. Perdew, Strongly Constrained and Appropriately Normed Semilocal Density Functional, Physical Review Letters, 2015,* ***115****, 036402.*
*53 A. Togo and I. Tanaka, First principles phonon calculations in materials science, Scripta Materialia, 2015,* ***108****, 1–5.*
*54 L. Bellaiche and D. Vanderbilt, Virtual crystal approximation revisited: Application to dielectric and piezoelectric properties of perovskites, Physical Review B, 2000,* ***61****, 7877–7882.*
*55 J. Zaanen, G. A. Sawatzky and J. W. Allen, Band gaps and electronic structure of transition-metal compounds, Physical Review Letters, 1985,* ***55****, 418–421.*
*56 J. Zaanen, G. A. Sawatzky and J. W. Allen, The electronic structure and band gaps in transition metal compounds, Journal of Magnetism and Magnetic Materials, 1986,* ***54–57****, 607–611.*
*57 https://www.ccryst.ehu.es*
*58 Y. Haven, The solubility of MgF2 in solid LiF, Recueil des Travaux Chimiques des Pays-Bas, 1950,* ***69****, 1505–1518.*
*59 X. Liu, S. K. Pandey and J. Feng, Silver(II) route to unconventional superconductivity, Physical Review B, 2022,* ***105****, 134519.*
*60 D. Rybicki, M. Jurkutat, S. Reichardt, C. Kapusta and J. Haase, Perspective on the phase diagram of cuprate high-temperature superconductors, Nature Communications, 2016,* ***7****, 11413.*
*61 N. Kowalski, S. S. Dash, P. Sémon, D. Sénéchal and A.-M. Tremblay, Oxygen hole content, charge-transfer gap, covalency, and cuprate superconductivity, Proceedings of the National Academy of Sciences, 2021,* ***118****, e2106476118.*

# Electronic Supplementary Information

## Exploring Intergrowth Structures with Flat [Ag(II)$F_2$] Layers to Mimic Oxocuprates(II): A Theoretical Study

Daniel Jezierski,*[a] Jose Lorenzana,*[b] and Wojciech Grochala*[a]

[a]*Center of New Technologies, University of Warsaw, 02089 Warsaw, Poland*

[b] *Institute for Complex Systems (ISC), CNR, 00185 Rome, Italy*

# I. Structural and magnetic data of [$AgF_2$] layers in intergrowth compounds.

Table SI1. Structural and magnetic data for [$AgF_2$] layers in *Pnma* structures employing DFT+U ($U_{Ag}$ = 5 eV) method.

| Near to the $AgF_2$ layer | Short name | Formula | dAg-Ag (A) [Å] | dAg-F ax. (C) [Å] | dAg-F eq. (B1) [Å] | dAg-F eq. (B2) [Å] | Ag-F-Ag angle α [°] | B1/B2 | $J_{2D}$ [meV] | | |
|---|---|---|---|---|---|---|---|---|---|---|---|
| | | | | | | | | | DFT+U | SCAN | HSE06 |
| 2xCs | Cs3 | $(CsMgF_3)_2CsAgF_3$ | 4.18 | 2.49 | 2.09 | 2.09 | 179.80 | 1.00 | -185.5 | -212.3 | |
| 2xCs | Cs2Rb | $(CsMgF_3)_2RbAgF_3$ | 4.15 | 2.51 | 2.08 | 2.08 | 179.80 | 1.00 | -201.2 | -237.7 | |
| 2xRb | CsRb2 | $(RbMgF_3)_2CsAgF_3$ | 4.14 | 2.36 | 2.02 | 2.12 | 175.80 | 0.95 | -139.6 | -165.3 | |
| 2xCs | Cs2K | $(CsMgF_3)_2KAgF_3$ | 4.13 | 2.52 | 2.07 | 2.07 | 179.60 | 1.00 | -211.0 | -255.5 | -220.6 |
| 2xK | CsK2 | $(KMgF_3)_2CsAgF_3$ | 4.12 | 2.22 | 2.01 | 2.27 | 148.00 | 0.89 | -29.7 | +0.1 | |
| 2xRb | Rb3 | $(RbMgF_3)_2RbAgF_3$ | 4.08 | 2.42 | 2.06 | 2.06 | 163.60 | 1.00 | -187.0 | -212.9 | |
| 2xRb | Rb2K | $(RbMgF_3)_2KAgF_3$ | 4.05 | 2.44 | 2.06 | 2.06 | 159.20 | 1.00 | -145.7 | -216.2 | |
| 2xK | RbK2 | $(KMgF_3)_2RbAgF_3$ | 4.02 | 2.39 | 2.08 | 2.07 | 151.00 | 1.00 | -182.6 | -167.0 | |
| 2xK | K3 | $(KMgF_3)_2KAgF_3$ | 3.99 | 2.40 | 2.07 | 2.07 | 149.47 | 1.00 | -147.2 | -179.0 | |

Table SI2. Structural and magnetic data for [$AgF_2$] layers in *P2₁/c* structures employing DFT+U ($U_{Ag}$ = 5 eV) method.

| Near to the AgF2 layer | short name | Formula | d Ag-Ag (A) [Å] | d Ag-F ax. (C) [Å] | | d Ag-F eq. (B1) [Å] | | d Ag-F eq. (B2) [Å] | | Ag-F-Ag angle α [°] | | dE of $P2_1/c$ – *Pnma* [meV/Ag] | $J_{2D}$ [meV] | |
|---|---|---|---|---|---|---|---|---|---|---|---|---|---|---|
| | | | | *I layer* | *II layer* | *I layer* | *II layer* | *I layer* | *II layer* | *I layer* | *II layer* | | *I layer* | *II layer* |
| Cs, Rb | mCs2Rb | $(CsMgF_3)_2RbAgF_3$ | 4.17 | 2.51 | 2.32 | 2.08 | 2.01 | 2.08 | 2.16 | 179.87 | 173.80 | +179.8 | -178.1 | -92.7 |
| Rb, Cs | mCsRb2 | $(RbMgF_3)_2CsAgF_3$ | 4.12 | 2.40 | 2.55 | 2.06 | 2.06 | 2.07 | 2.06 | 169.54 | 179.92 | -183.7 | -180.2 | -207.2 |
| Rb, K | mI_RbCsK | $RbCsKMg_2AgF_9$ | 4.11 | 2.40 | 2.26 | 2.05 | 2.01 | 2.08 | 2.24 | 169.95 | 150.36 | | -144.7 | -145.3 |
| K, Cs | mII_RbCsK | $RbCsKMg_2AgF_9$ | 4.11 | 2.24 | 2.58 | 2.01 | 2.05 | 2.24 | 2.05 | 149.88 | 179.94 | | -24.6 | -169.3 |
| Cs, Rb | mIII_RbCsK | $RbCsKMg_2AgF_9$ | 4.09 | 2.57 | 2.42 | 2.05 | 2.06 | 2.05 | 2.07 | 179.64 | 164.34 | | -218.4 | -183.3 |
| K, Cs | mCsK2 | $(KMgF_3)_2CsAgF_3$ | 4.08 | 2.28 | 2.59 | 2.01 | 2.04 | 2.22 | 2.04 | 149.65 | 179.57 | -259.6 | -43.8 | -191.7 |
| Rb, K | mRb2K | $(RbMgF_3)_2KAgF_3$ | 4.06 | 2.44 | 2.32 | 2.06 | 2.02 | 2.06 | 2.19 | 160.84 | 149.97 | +167.2 | -179.6 | -76.7 |
| K, Rb | mRbK2 | $(KMgF_3)_2RbAgF_3$ | 4.02 | 2.39 | 2.47 | 2.08 | 2.06 | 2.08 | 2.06 | 151.26 | 156.55 | -168.6 | -148.4 | -175.2 |

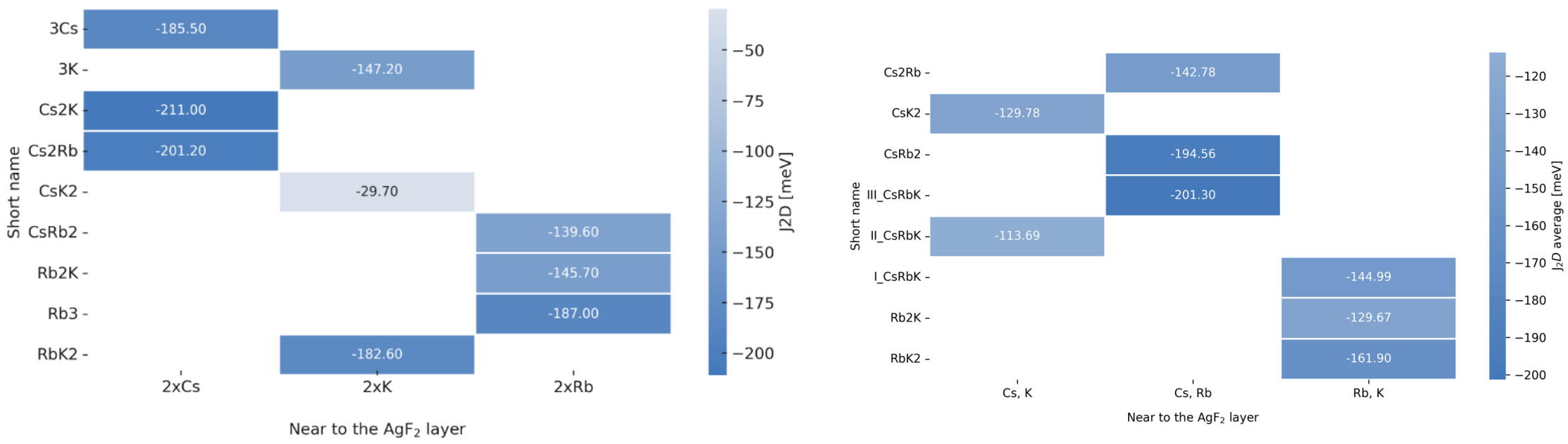

Figure SI1. $J_{2D}$ values for $[AgF_2]$ layer depending on the neighboring alkali metal in the *Pnma* (left) and $P2_1/c$ (right) structure for (DFT+U method).

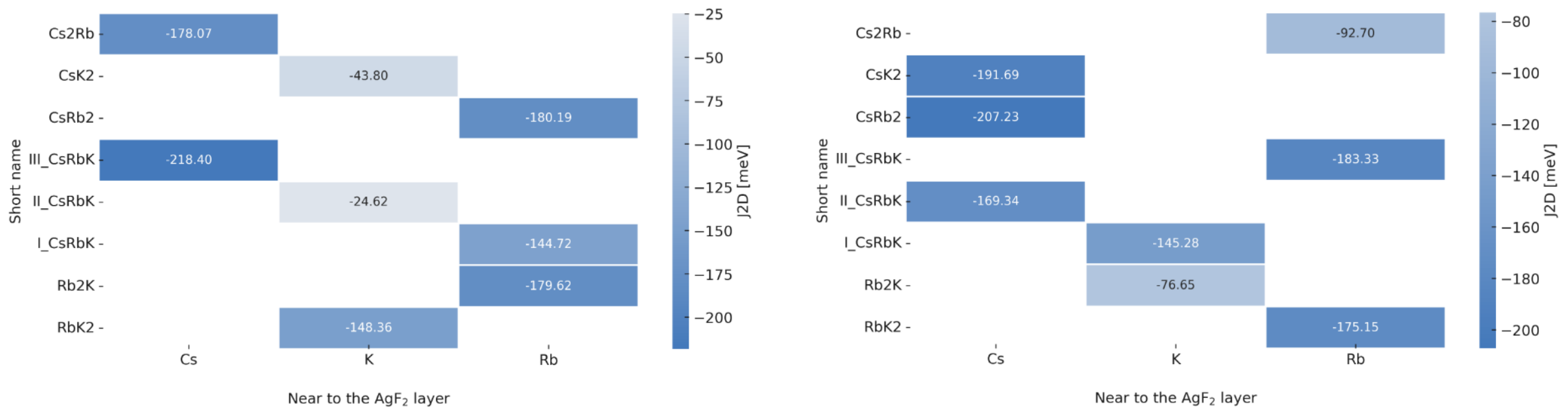

Figure SI2: $J_{2D}$ values of the I layer $[AgF_2]$ (left) and II layer $[AgF_2]$ (right) as a dependence of neighboring alkali metals in the $P2_1/c$ crystal structure (DFT+U method).

***Pnma***

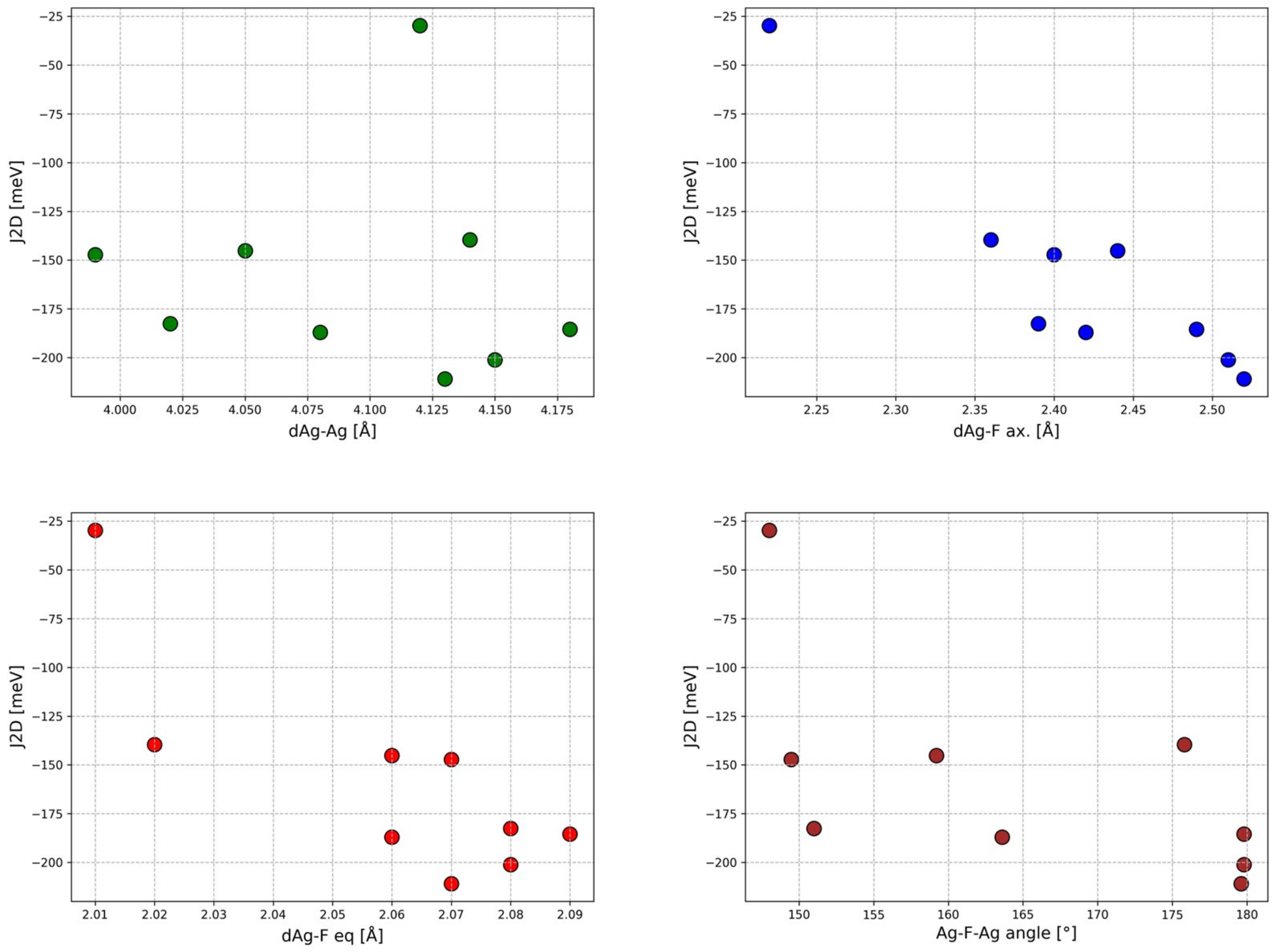

Figure SI3. $J_{2D}$ values of $[AgF_2]$ layers depending on the Ag-Ag, Ag-F axial and Ag-F equatorial distances and Ag-F-Ag bond angles for *Pnma* phases.

***P2₁/c***

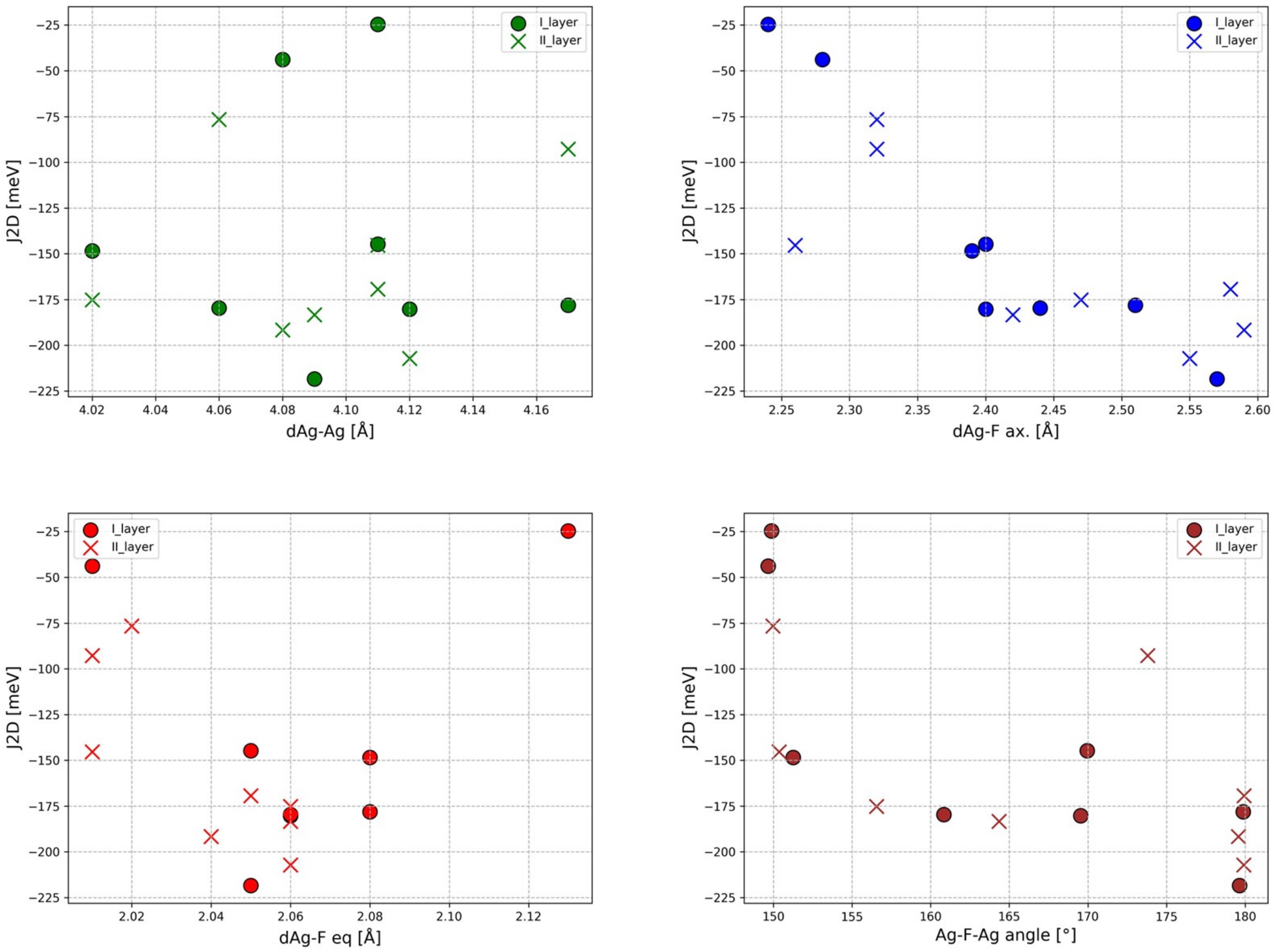

Figure SI4. $J_{2D}$ values of $[AgF_2]$ layers depending on the Ag-Ag, Ag-F axial and Ag-F equatorial distances and Ag-F-Ag bond angles for $P2_1/c$ phases.

## II. Reactions considered and their energy effects.

For $(XMgF_3)_2XAgF3$ energy effect was calculated as follows (for X = Rb, K or Cs):

$$dE_1 = E_{(XMgF_3)_2XAgF_3} - (2E_{MMgF_3} + E_{XAgF_3})$$

$$dE_3 = E_{(XMgF_3)_2XAgF_3} - (E_{X_2AgF_4} + E_{XMgF_3} + E_{MgF_2})$$

$$dE_4 = E_{(XMgF_3)_2XAgF_3} - (E_{X_2MgF_4} + E_{XMgF_3} + E_{AgF_2})$$

For $(XMgF_3)_2YAgF_3$ energy effect was calculated as follows (for X = Rb, K or Cs and Y = Rb, K, Cs):

$$dE_1 = E_{(XMgF_3)_2YAgF_3} - (2E_{XMgF_3} + E_{YAgF_3})$$

$$dE_2 = E_{(XMgF_3)_2YAgF_3} - (E_{XMgF_3} + E_{YMgF_3} + E_{XAgF_3})$$

$$dE_3 = E_{(XMgF_3)_2YAgF_3} - (E_{X_2AgF_4} + E_{YMgF_3} + E_{MgF_2})$$

$$dE_4 = E_{(XMgF_3)_2YAgF_3} - (E_{X_2MgF_4} + E_{YMgF_3} + E_{AgF_2})$$

If X =Rb and Y = K as in $(RbMgF_3)_2KAgF_3$ compound, the energy effects were calculated as follows:

$$dE_1 = E_{(RbMgF_3)_2KAgF_3} - (2E_{RbMgF_3} + E_{KAgF_3})$$

$$dE_2 = E_{(RbMgF_3)_2KAgF_3} - (E_{RbMgF_3} + E_{KMgF_3} + E_{RbAgF_3})$$

$$dE_3 = E_{(RbMgF_3)_2KAgF_3} - (E_{Rb_2AgF_4} + E_{KMgF_3} + E_{MgF_2})$$

$$dE_4 = E_{(RbMgF_3)_2KAgF_3} - (E_{Rb_2MgF_4} + E_{KMgF_3} + E_{AgF_2})$$

For $XYZMg_2F_7AgF_2$ phases energy effect was calculated as follows (as example for X= Rb, Y = K, Z = Cs):

$$dE_1 = E_{RbCsKMg_2AgF_9} - (E_{ZMgF_3} + E_{XAgF_3} + E_{YMgF_3})$$

$$dE_2 = E_{RbCsKMg_2AgF_9} - (E_{ZMgF_3} + E_{XMgF_3} + E_{YAgF_3})$$

$$dE_3 = E_{RbCsKMg_2AgF_9} - (E_{ZAgF_3} + E_{XMgF_3} + E_{YMgF_3})$$

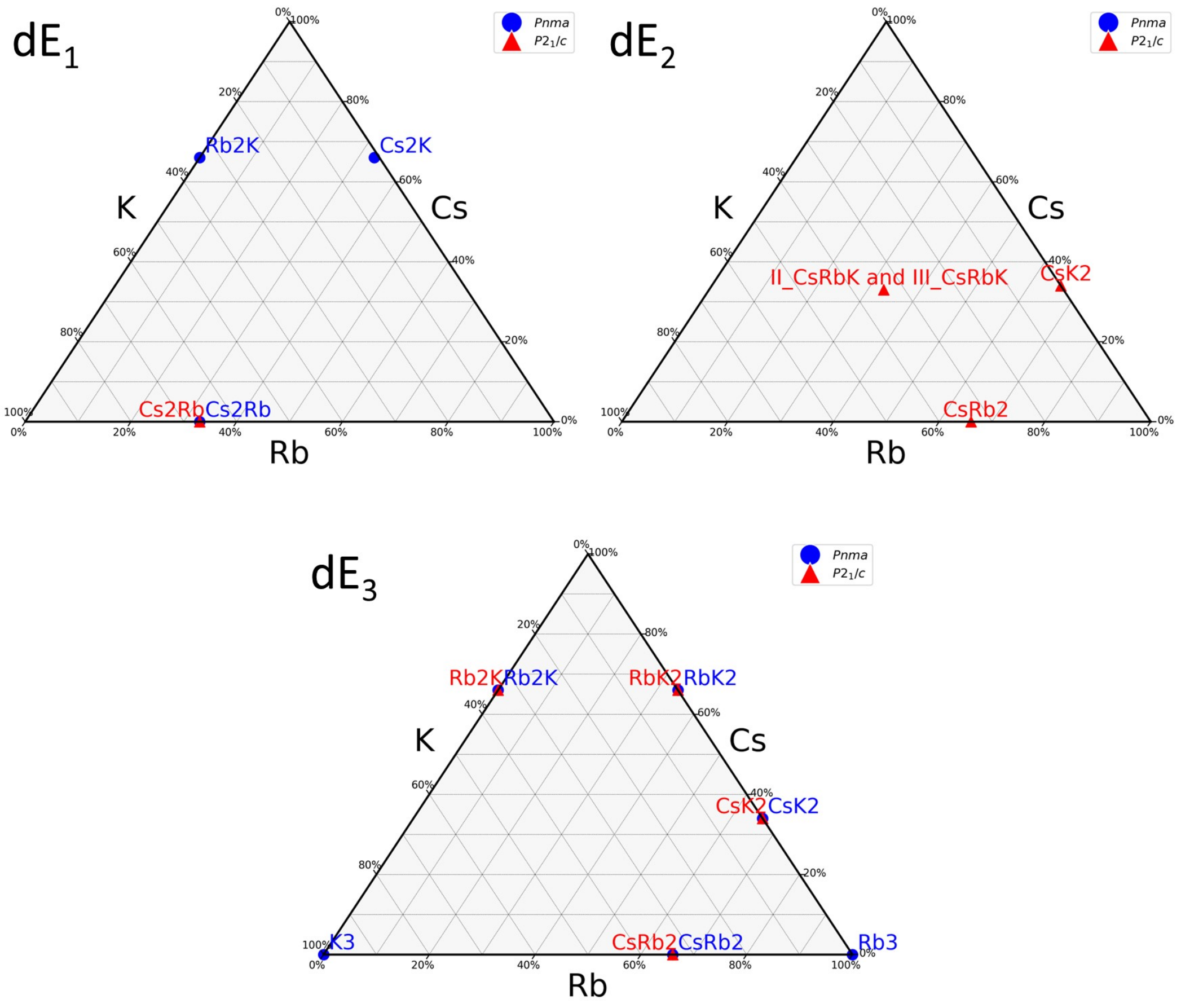

Figure SI5. Ternary phase diagram Cs-K-Rb for which $dE_1$, $dE_2$ and $dE_3$ are negative.

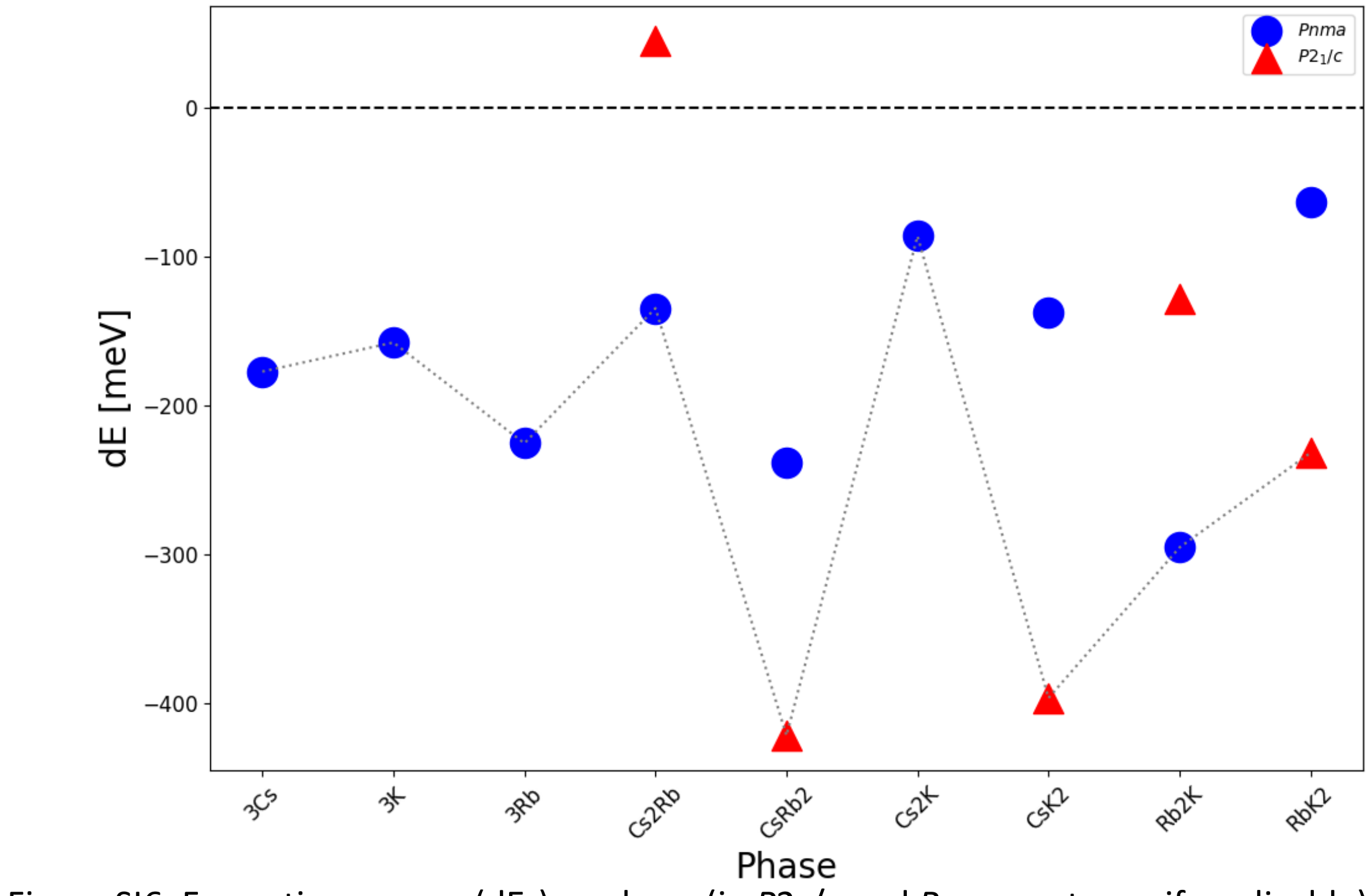

Figure SI6. Formation energy ($dE_4$) vs phase (in $P2_1/c$ and *Pnma* systems, if applicable)

## III. Density of states for selected compounds – magnetic and non-magnetic solutions. Electronic band structure for Cs2K.

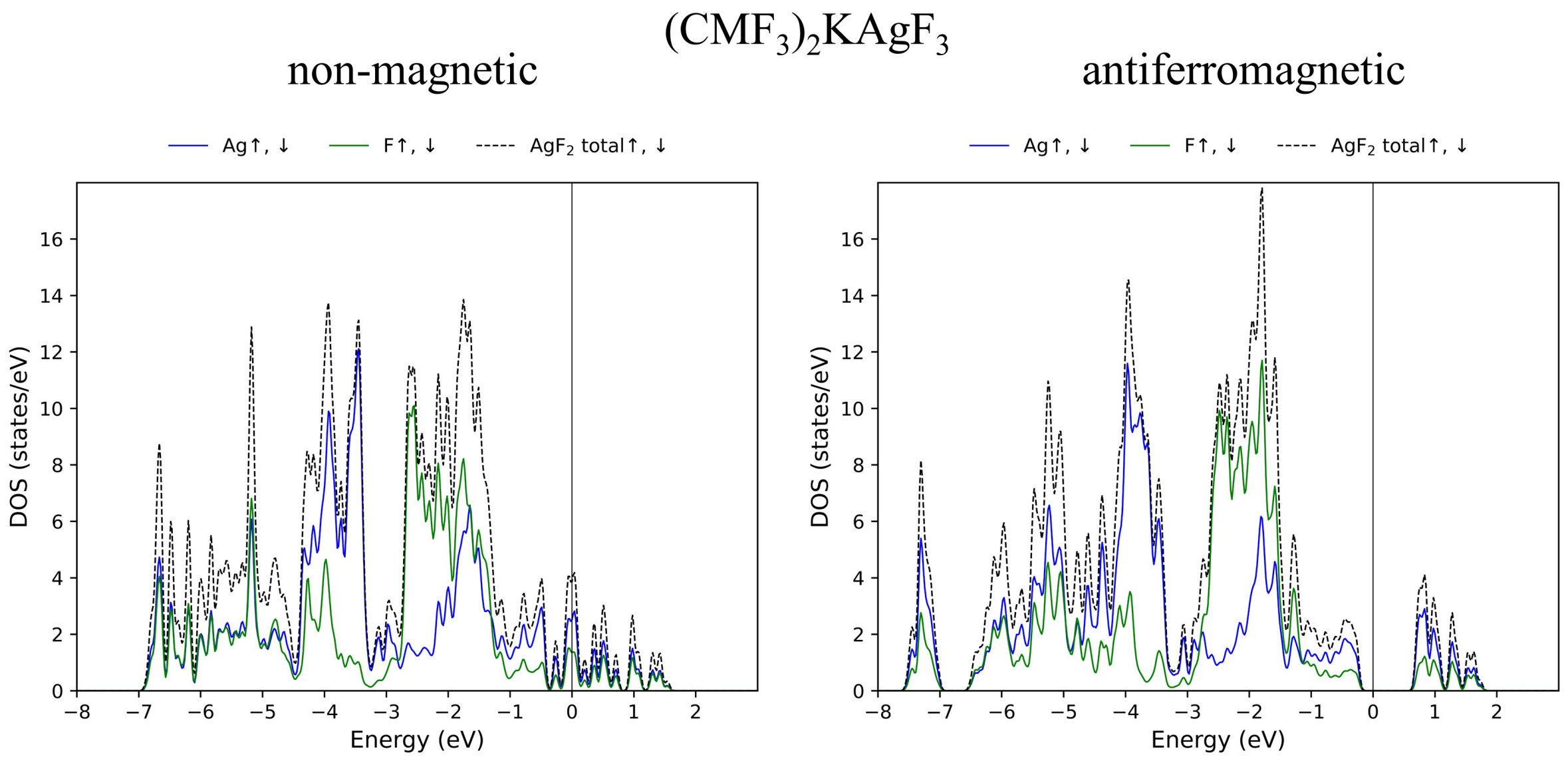

Figure SI7. DOS for Cs2K without spin-polarization (non-magnetic solution) on left and in antiferromagnetic solution on right. Only Ag and F states are presented.

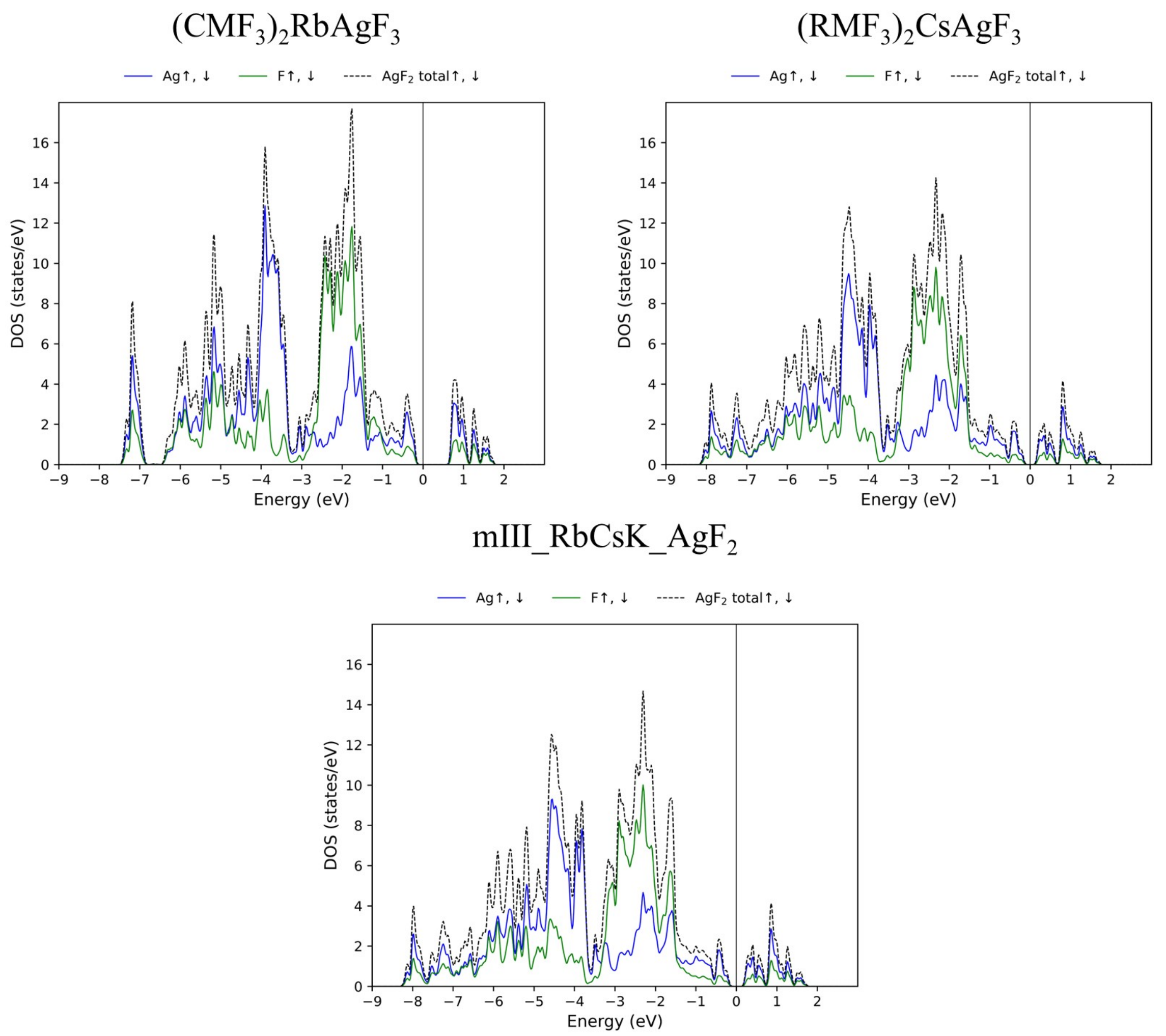

Figure SI8. DOS for **Cs2Rb, Rb2Cs** and **mIII_RbCsK** in antiferromagnetic solution. Only Ag and F states are presented.

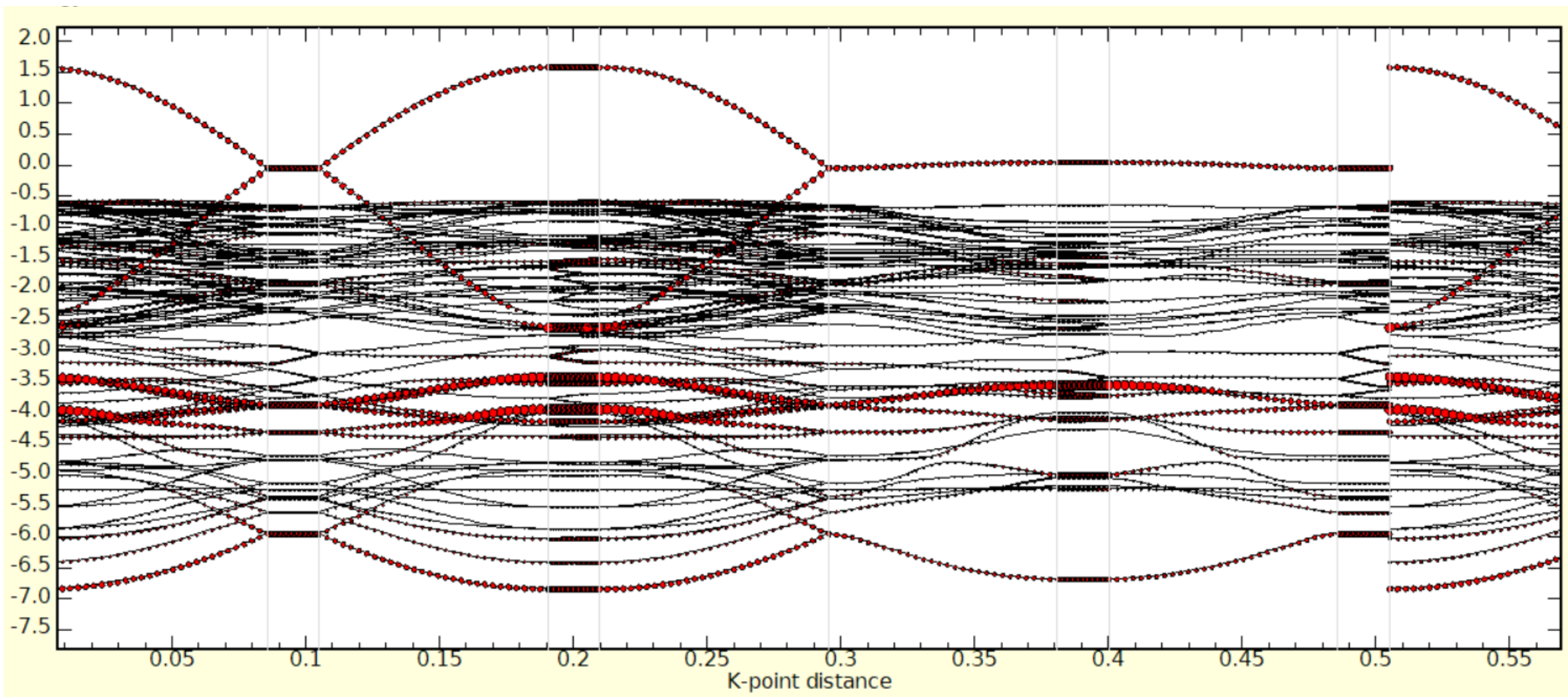

Figure SI9. Band structure for Non-magnetic solution of **Cs2K** (*Pnma* phase). Red dots correspond to the contribution from Ag*d* states from [$AgF_2$] layer.

# IV. Doping of Cs2K.

Table SI3 Structural, magnetic, and electronic parameters for Cs2K compounds in scenarios of gradual replacement of $K^+$ by $Ba^{2+}$. **A** for Ag-Ag distance, **C** for Ag-F apical bond, **B1** and **B2** for Ag-F equatorial bonds and **α** for Ag-F-Ag angle, **D** for distance between [$AgF_2$] layer to $M^+/M^{2+}$ cation. UHB is the width of the Upper Hubbard Band. Doping level, $\delta n_{Ag(UHB)}$, refers to the populated part of UHB.

| % Ba | a=c [Å] | b [Å] | V [Å³] | D [Å] | A [Å] | C [Å] | B1=B2 [Å] | α [°] | Ag mag. mom. [$\mu_B$] | Doping level, $\delta n_{Ag(UHB)}$ [%] | UHB [eV] |
|---|---|---|---|---|---|---|---|---|---|---|---|
| **0** | 5.842 | 25.975 | 886.312 | 7.12 | 4.13 | 2.53 | 2.07 | 179.60 | 0.48 | 0.0 | 1.242 |
| **1** | 5.839 | 26.053 | 888.176 | 7.14 | 4.13 | 2.54 | 2.06 | 179.97 | 0.48 | 6.0 | 1.262 |
| **2.5** | 5.839 | 26.083 | 889.344 | 7.14 | 4.13 | 2.55 | 2.06 | 179.96 | 0.47 | 6.7 | 1.301 |
| **5** | 5.840 | 26.136 | 891.498 | 7.16 | 4.13 | 2.57 | 2.06 | 179.93 | 0.46 | 12.1 | 1.319 |
| **10** | 5.842 | 26.244 | 895.763 | 7.18 | 4.13 | 2.61 | 2.07 | 179.90 | 0.42 | 13.9 | 1.377 |
| **15** | 5.845 | 26.351 | 900.218 | 7.21 | 4.13 | 2.65 | 2.07 | 179.86 | 0.38 | 13.7 | 1.435 |
| **20** | 5.849 | 26.445 | 904.592 | 7.23 | 4.14 | 2.68 | 2.07 | 179.89 | 0.34 | 18.4 | 1.515 |
| **25** | 5.854 | 26.532 | 909.206 | 7.25 | 4.14 | 2.71 | 2.07 | 179.91 | 0.29 | 22.8 | 1.592 |

Doping level is calculated following formula below. Therefore it is the part (%) of UHB band occupied, as a result of $K^+$ by $Ba^{2+}$ replacement in the Cs2K structure.

$$\delta n_{Ag(UHB)}\% = \frac{\int_{E_0(UHB)}^{E_F(UHB)} (eDOS)dE}{\int_{E_0(UHB)}^{E_1(UHB)} (eDOS)dE} * 100\%$$

$E_0$ – lower part of UHB, $E_F$ – Fermi level, $E_1$ – upper part of UHB.

Table SI4 Structural, magnetic, and electronic parameters for Cs2K compounds in scenarios of gradual replacement of $Mg^{2+}$ by $Li^+$. **A** for Ag-Ag distance, **C** for Ag-F apical bond, **B1** and **B2** for Ag-F equatorial bonds and **α** for Ag-F-Ag angle, **D** for distance between [$AgF_2$] layer to [$Li^+/Mg^{2+}F_2$] layer. Doping level, $\delta n_{Ag(\beta-dx^2-y^2)}$, $\delta n_F$, for depopulation level of apical fluorine atoms $p_{(z)}$ and , $\delta n_{Ag(dz^2)}$ – depopulation level of Ag$d_{(z^2)}$ states.

| % Li | a=c [Å] | b [Å] | V [Å³] | D [Å] | A [Å] | C [Å] | B1=B2 [Å] | α [°] | Ag mag. mom. [$\mu_B$] | F apical mag. mom [$\mu_B$] | Doping level [%] | | |
|---|---|---|---|---|---|---|---|---|---|---|---|---|---|
| | | | | | | | | | | | $\delta n_{Ag(\beta-dx^2-y^2)}$ | $\delta n_{Ag(dz^2)}$ | $\delta n_F$ |
| **0** | 5.842 | 25.975 | 886.312 | 4.54 | 4.13 | 2.53 | 2.07 | 179.60 | 0.48 | 0.00 | 0 | 0 | 0 |
| **1** | 5.838 | 26.037 | 887.462 | 4.55 | 4.13 | 2.53 | 2.06 | 179.89 | 0.48 | 0.01 | 0.2 | 0.5 | 0.2 |
| **2.5** | 5.839 | 26.026 | 887.441 | 4.55 | 4.13 | 2.53 | 2.06 | 179.94 | 0.48 | 0.01 | 1.1 | 1.6 | 0.5 |
| **5** | 5.845 | 25.924 | 885.810 | 4.52 | 4.13 | 2.49 | 2.07 | 179.99 | 0.48 | 0.00 | 1.1 | 3.6 | 7.9 |
| **10** | 5.857 | 25.715 | 882.229 | 4.47 | 4.14 | 2.40 | 2.07 | 180.00 | 0.47 | 0.01 | 1.4 | 4.4 | 7.4 |
| **15** | 5.866 | 25.602 | 880.898 | 4.45 | 4.15 | 2.34 | 2.07 | 180.00 | 0.47 | 0.02 | 1.4 | 5.7 | 6.7 |
| **20** | 5.873 | 25.543 | 880.994 | 4.45 | 4.15 | 2.26 | 2.08 | 180.00 | 0.44 | 0.03 | 1.3 | 11.0 | 6.6 |
| **25** | 5.923 | 25.417 | 891.813 | 4.44 | 4.19 | 2.07 | 2.09 | 180.00 | 0.51 | 0.12 | * | * | * |

* rearrangement of geometry, consideration not applicable

$$\delta n_{Ag(\beta-dx^2-y^2)}\% = \frac{\int_{E_F(LHB)}^{E_1(LHB)} (eDOS)dE}{\int_{E_0(LHB)}^{E_1(LHB)} (eDOS)dE} * 100\%$$

$E_0$ – lower part of LHB, $E_F$ – Fermi level, $E_1$ – upper part of LHB. This formula was applied to calculate depopulation level of β– $d_{(x2\text{-}y2)}$ of silver, and also to calculate depopulation level of F$p_{(z)}$ and Ag$d_{(z2)}$ orbitals.

*Electronic density of states for 25% of Li doping level.*

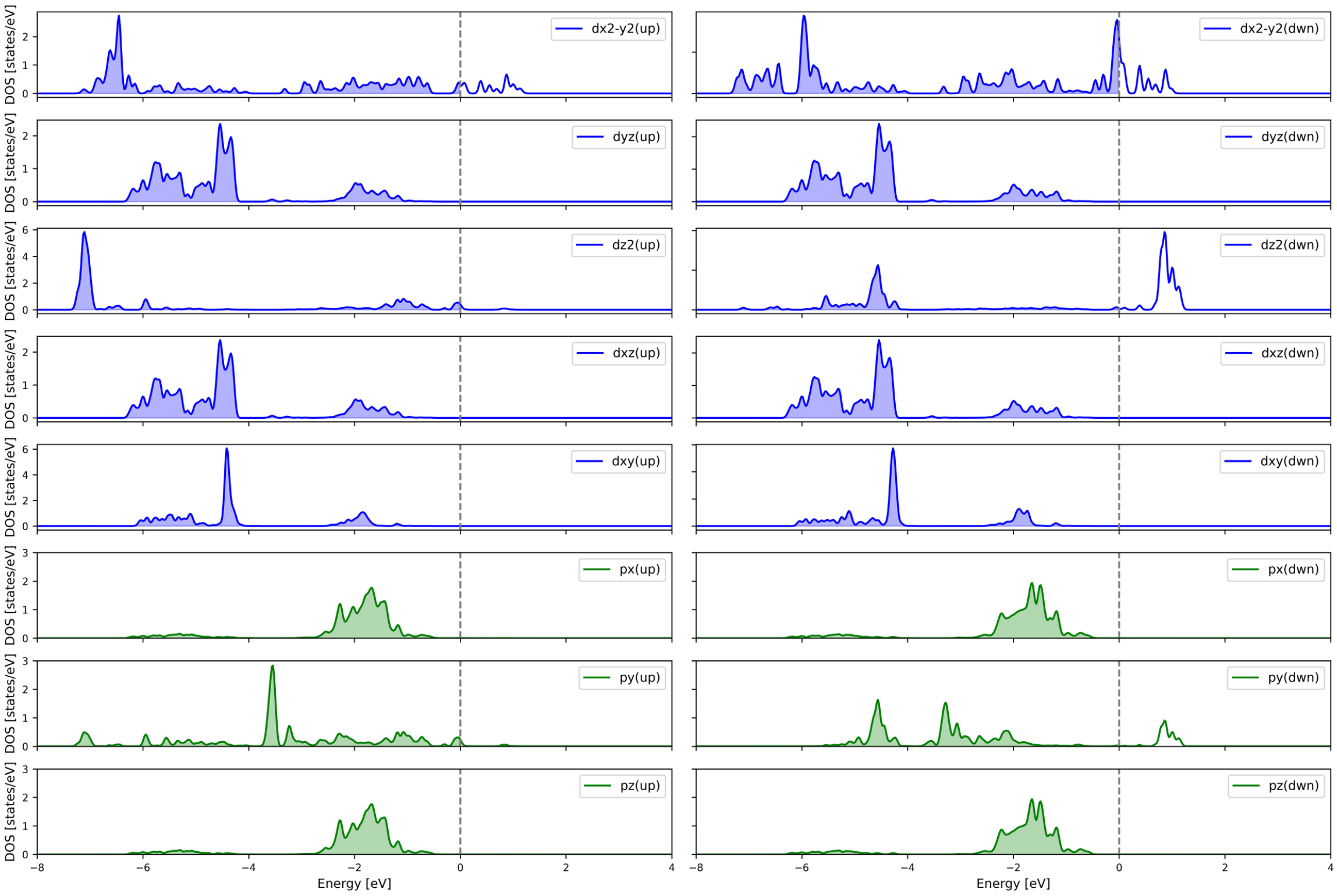

Figure SI10. Electronic density of states $AgF_2$ for 25% Li doping level. Ag states in blue, F states in green

## V. Hole doping of LCO

Table SI4 Structural, magnetic, and electronic parameters for LCO ($La_2CuO_4$) compounds in scenarios of gradual replacement of $La^{3+}$ by $Ba^{2+}$. **A** for Cu-Cu distance, **C** for Cu-O apical bond, **B1** and **B2** for Cu-O equatorial bonds and **α** for Cu-O-Cu angle. Doping level, $\delta n_{Cu(LHB)}$, stems for depopulation of Lower Hubbard Band (LHB), $\delta n_O$, for depopulation level of apical oxygen and , $\delta n_{Cu(dz^2)}$ – depopulation level of Cu$d_{(z^2)}$ states. The doping level was calculated in the same manner as in the previous chapter.

| % Ba | a [Å] | b [Å] | c [Å] | V [Å³] | A [Å] | C [Å] | B1=B2 [Å] | α [°] | Cu mag. mom [μB] | F apical mag. mom [μB] | Doping level [%] | | |
|---|---|---|---|---|---|---|---|---|---|---|---|---|---|
| | | | | | | | | | | | $\delta n_{Cu(LHB)}$ | $\delta n_{Cu(dz^2)}$ | $\delta n_O$ |
| **0** | 5.271 | 5.408 | 12.994 | 370.420 | 3.78 | 2.42 | 1.89 | 172.35 | 0.52 | 0.01 | 0 | 0 | 0 |
| **5** | 5.290 | 5.360 | 13.081 | 370.883 | 3.77 | 2.39 | 1.88 | 174.70 | 0.52 | 0.01 | 1.4 | 4.7 | 7.3 |
| **10** | 5.314 | 5.317 | 13.142 | 371.320 | 3.76 | 2.36 | 1.88 | 179.44 | 0.51 | 0.02 | 1.8 | 6.3 | 7.9 |
| **15** | 5.325 | 5.327 | 13.162 | 373.370 | 3.77 | 2.34 | 1.88 | 179.99 | 0.51 | 0.04 | 1.7 | 6.7 | 8.4 |

## VI. Cif file for Cs2K.

### Cs2K

```
     5.8406977654       0.0000000000       0.0000000000
     0.0000000000      25.9749107361       0.0000000000
     0.0000000000       0.0000000000       5.8420829773
  Ag   Mg   Cs    K    F
   4    8    8    4   36
Direct
   0.000000000      0.000000000      0.000000000
   0.500000000      0.000000000      0.500000000
   0.000000000      0.500000000      0.000000000
   0.500000000      0.500000000      0.500000000
   0.000020000      0.825380027      0.000000000
   0.999979973      0.174619973      0.000000000
   0.499980003      0.174619973      0.500000000
   0.500020027      0.825380027      0.500000000
   0.999979973      0.325380027      0.000000000
   0.000020000      0.674619973      0.000000000
   0.500020027      0.674619973      0.500000000
   0.499980003      0.325380027      0.500000000
   0.000060000      0.908309996      0.500020027
   0.999939978      0.091689996      0.499979973
   0.499940008      0.091689996      0.000020027
   0.500060022      0.908309996      0.999979973
   0.999939978      0.408309996      0.499979973
   0.000060000      0.591690004      0.500020027
   0.500060022      0.591690004      0.999979973
   0.499940008      0.408309996      0.000020027
   0.999960005      0.250000000      0.500140011
   0.000040000      0.750000000      0.499859989
   0.500039995      0.750000000      0.000140011
   0.499960005      0.250000000      0.999859989
   0.750039995      0.824339986      0.249970004
   0.249960005      0.175660014      0.750029981
   0.749960005      0.175660014      0.749970019
   0.250039995      0.824339986      0.250029981
   0.249960005      0.324339986      0.750029981
   0.750039995      0.675660014      0.249970004
   0.250039995      0.675660014      0.250029981
   0.749960005      0.324339986      0.749970019
   0.250050008      0.675759971      0.749979973
   0.749949992      0.324240029      0.250020027
   0.249949992      0.324240029      0.249979973
   0.750050008      0.675759971      0.750020027
   0.749949992      0.175759971      0.250020027
   0.250050008      0.824240029      0.749979973
   0.750050008      0.824240029      0.750020027
   0.249949992      0.175759971      0.249979973
   0.749220014      0.999849975      0.249219999
   0.250779986      0.000150000      0.750779986
```

| | | |
|---|---|---|
| 0.750779986 | 0.000150000 | 0.749220014 |
| 0.249220014 | 0.999849975 | 0.250779986 |
| 0.250779986 | 0.499850005 | 0.750779986 |
| 0.749220014 | 0.500150025 | 0.249219999 |
| 0.249220014 | 0.500150025 | 0.250779986 |
| 0.750779986 | 0.499850005 | 0.749220014 |
| 0.500029981 | 0.902700007 | 0.500199974 |
| 0.499970019 | 0.097300000 | 0.499800026 |
| 0.999970019 | 0.097300000 | 0.000199974 |
| 0.000029981 | 0.902700007 | 0.999800026 |
| 0.499970019 | 0.402700007 | 0.499800026 |
| 0.500029981 | 0.597299993 | 0.500199974 |
| 0.000029981 | 0.597299993 | 0.999800026 |
| 0.999970019 | 0.402700007 | 0.000199974 |
| 0.499960005 | 0.250000000 | 0.500419974 |
| 0.500039995 | 0.750000000 | 0.499580026 |
| 0.000039995 | 0.750000000 | 0.000419974 |
| 0.999960005 | 0.250000000 | 0.999580026 |